\documentclass[12pt,preprint]{aastex}
\begin{document}
\title{Multiwaveband Polarimetric Observations of 15 Active Galactic Nuclei at High
Frequencies: Correlated Polarization Behavior}
\author{Svetlana G. Jorstad\altaffilmark{1,2}, Alan P. Marscher\altaffilmark{1},
Jason A. Stevens\altaffilmark{3}, Paul S. Smith\altaffilmark{4}, 
James R. Forster\altaffilmark{5}, Walter K. Gear\altaffilmark{6}, 
Timothy V. Cawthorne \altaffilmark{7}, Matthew L. Lister\altaffilmark{8}, 
Alastair M. Stirling\altaffilmark{9}, Jos\'e L. G\'omez\altaffilmark{10}, 
Jane S. Greaves\altaffilmark{11}, and E. Ian Robson\altaffilmark{12}}
\altaffiltext{1}{Institute for Astrophysical Research, Boston University,
725 Commonwealth Ave., Boston, MA 02215-1401; jorstad@bu.edu, marscher@bu.edu}
\altaffiltext{2}{Sobolev Astronomical Institute, St. Petersburg State University,
Universitetskij pr. 28, 198504 St. Petersburg, Russia}
\altaffiltext{3}{Centre for Astrophysics Research, Science and Technology Centre,
University of Hertfordshire, College Lane, Herts AL10 9AB, UK; J.A.Stevens@herts.ac.uk}
\altaffiltext{4}{Steward Observatory, The University of Arizona, Tucson, AZ 85721;
psmith@as.arizona.edu}
\altaffiltext{5}{Hat Creek Observatory, University of California, Berkeley, 42231 
Bidwell Rd. Hatcreek, CA 96040; rforster@astro.berkeley.edu}
\altaffiltext{6}{School of Physics and Astronomy, Cardiff University, 5, The Parade
 Cardiff CF2 3YB, Wales, UK; Walter.Gear@astro.cf.ac.uk}
\altaffiltext{7}{Center for Astrophysics, University of Central Lancashire, Preston,
PR1 2HE, UK; tvcawthorne@uclan.ac.uk}
\altaffiltext{8}{Department of Physics, Purdue University, 525 Northwestern Ave.,
West Lafayette, IN 47907-2036; mlister@physics.purdue.edu}
\altaffiltext{9}{University of Manchester, Jodrell Bank Observatory, Macclesfield,
Cheshire, SK11 9DL, UK (current address); ams@jb.man.ac.uk}
\altaffiltext{10}{Insituto de Astrof\'{\i}sica de Andaluc\'{\i}a (CSIC), Apartado 3004,
Granada 18080, Spain; jlgomez@iaa.es}
\altaffiltext{11}{School of Physics and Astronomy, University of St Andrews, North Haugh, St Andrews, Fife KY16 9SS, UK; jsg5@st-and.ac.uk} 
\altaffiltext{12}{Astronomy Technology Centre, Royal Observatory, Blackford Hill, 
Edinburgh EH9 3HJ, UK; eir@roe.ac.uk}
\shorttitle{Multifrequency Polarization Studying}
\shortauthors{Jorstad et al.}
\begin{abstract}
We report on multi-frequency linear polarization monitoring of 15 active galactic nuclei containing highly relativistic jets with apparent speeds from $\sim$4~$c$ to $>$40$c$. The measurements were obtained at optical, 1~mm, and 3~mm
wavelengths, and at 7~mm with the Very Long Baseline Array. 
The data show a wide range in degree of linear polarization among the sources, 
from $<$1\% to $>$30\%, and interday polarization variability
in individual sources.
The polarization properties suggest separation of the sample
into three groups with low, intermediate, and high variability 
of polarization in the core at 7~mm : LVP, IVP, and HVP, respectively.
The groups are partially associated with the common classification of active 
galactic nuclei as radio galaxies and quasars with low optical polarization
(LVP), BL Lacertae objects (IVP), and highly optically polarized quasars (HVP).
Our study investigates correlations between total flux, fractional 
polarization, and polarization position angle at the different wavelengths. 
We interpret the polarization properties 
of the sources in the sample through models in which weak shocks compress
turbulent plasma in the jet. The differences in the
orientation of sources with respect to the observer, jet kinematics,
and abundance of thermal matter external to the jet near the core
can account for the diversity in the polarization properties.
The results provide strong evidence that the optical polarized
emission originates in shocks, most likely situated between 
the 3~mm and 7~mm VLBI cores. They also support the idea that the 1~mm core lies 
at the edge of the transition zone between electromagnetically 
dominated and turbulent hydrodynamical sections of the jet. 
\end{abstract}


\section{Introduction}
Magnetic fields play a prominent role in the physical processes that occur in the jets
of active galactic nuclei (AGN). The leading model for jet production, acceleration,
and collimation involves poloidal
magnetic fields that are wound up by the differential rotation of a rotating disk or 
ergosphere surrounding a central supermassive black hole \citep[e.g.,][]{McK06,MKU00}.
The twisted field propagates outward as Poynting flux in the polar directions, with
eventual conversion into a well-focused relativistic plasma flow
\citep[e.g.,][]{VK04,MN06}. Within this zone, the magnetic field should maintain a
tight helical pattern.

Beyond the jet acceleration region---which may extend over hundreds or thousands of
gravitational radii from the black hole \citep{VK04,MAR06}---the jet may become turbulent
or subject to velocity shear. In the former case, the magnetic field should be chaotic,
with any line of sight passing through many turbulent cells and significant differences
in both strength and direction of the field in adjacent cells. In contrast, velocity shear
stretches and orders the field lines along the flow \citep[e.g.,][]{Laing80}.
Shock waves passing through the flow (or vice versa) will compress the component of
the field that is parallel to the shock front, which imposes order even on a magnetic
field that is completely chaotic in front of the shock.

The geometry and degree of order of the magnetic field are therefore key indicators
of the physical conditions in a jet. Because the primary emission mechanism at
radio to optical wavelengths is synchrotron radiation, the linear
polarization of the continuum can be used as a probe of the magnetic field. 
We can also use the fractional
polarization and direction of the electric vector position angle (EVPA) to
identify distinct features in the jet that are observed at different wavelengths.
The most prominent features on VLBI images of jets in radio-loud AGN are (1) the core, which is
the bright,
very compact section at the narrow end of a one-sided jet, and (2) condensations in the
flow that appear as bright knots, often called ``components'' of the jet. 
The core likely lies
some distance from the central engine of the AGN, probably either near the end or beyond
the zone of acceleration and collimation of the jet \citep{MAR06}. The knots usually
separate from the core at apparent superluminal speeds, but roughly stationary knots
are also present in many jets, perhaps representing standing shocks
\citep[e.g.,][hereafter J05]{J01,KL04,LIST06,J05}. The apparent speeds
of the components can exceed
$\beta_{\rm app} \sim 40c$ (J05) that requires the Lorentz factor of
the flow exceeds $\beta_{\rm app}$. The leading model identifies the moving knots
as propagating shocks, either transverse to the jet axis \citep[e.g.,][]{HAA85,HAA89,MG85}
or at an oblique angle \citep{PH05}. Polarization studies can help to determine
whether this model is viable.
 
The standard paradigm of transverse shocks propagating down a relativistic jet
can explain many aspects of the time variability of the
brightness, polarization, and structure at radio wavelengths
\citep[e.g., ][]{HAA85, HAA89, CW88, W95}. However,
this model makes predictions---e.g., that the magnetic field of a knot 
should be transverse 
to jet axis---that often do not match observations.
Another possibility is that there is a systematically ordered component of the magnetic 
field, for example one with a helical geometry \citep[e.g., ][]{LPG05}, in the jet that
modulates its brightness and polarization variability. In fact, a helical magnetic field is 
required in the magnetohydrodynamical jet launching models \citep{MKU00,VK04}.
\citet{GAB06} provides some support for such a geometry in recent low frequency VLBI
observations of Faraday rotation in AGN jets. However, this structure can be related to the magnetic field in the Faraday screen surrounding the jet since 
high frequency radio polarization mapping reveals aspects that are difficult to 
explain by a helical field \citep{PH05,ZT05}.

The situation is even more confusing in 
the optical and near-IR regions, where variations in flux and polarization are  often extremely 
rapid, and sub-milliarcsecond resolution is not available
\citep[e.g., ][]{HT80,MSW87,SMITH87,MEAD90}. 
Furthermore, existing optical data indicate that the connection between variations in
brightness and polarization is tenuous \citep[e.g.,][]{SMITH96}. These factors complicate
interpretation of the optical/near-IR emission, since detailed models are needed but the
data are not extensive enough to guide them. An alternative approach is to investigate the 
global polarization behavior across a sizable range of the electromagnetic spectrum.
Such studies \citep{WILLS92,GSS96,LS00,GRSS06} have produced convincing evidence 
for a connection between the radio and optical polarization properties of AGN, 
suggestive of a common, probably cospatial, origin for the emission at
these two wavebands. However, such studies have been based on single-epoch
measurements, whereas the investigation we present here involves multi-epoch observations
of fifteen objects. This allows us to use both the time and frequency domains to explore the
geometry and degree of ordering of the magnetic field and other properties of the
emission regions in relativistic jets.

Bright jets with high apparent
superluminal motion are a prevalent feature of blazars, a classification that includes
BL~Lac objects and optically violently variable quasars (OVVs) that are quite rare in the
general AGN population \citep{KL04,LIST06}. The prominence of the jets from radio to
optical wavelengths and the pronounced variability of their polarized emission makes
blazars and radio galaxies with blazar-like behavior ideal objects for probing the magnetic
fields in jets through multi-epoch polarization studies.
 
We have carried out a 3-yr monitoring program
of fifteen radio-loud, highly variable AGN. The program combines roughly bimonthly, high resolution polarized and total intensity
radio images with optical, submillimeter-wave (sub-mm), and millimeter-wave (mm) 
polarization observations performed at many of the same epochs. The sample includes 
objects that are usually brighter than about 2~Jy at 7~mm and 1~Jy at 1~mm, with a mixture of 
quasars (0420$-$014, 0528+134, 3C~273, 3C~279, 1510$-$089,
3C~345, CTA~102, and 3C~454.3), BL~Lac objects (3C~66A, OJ~287, 1803+784, 1823+568,
and BL~Lac), and radio galaxies with bright compact radio jets (3C~111 and 3C~120).
In J05, we investigated the properties of the jets on milliarcsecond 
scales and determined the apparent velocities of more than 100 jet features. For 
the majority of these components, we derived Doppler factors using a new method based 
on comparison of timescale of decline in flux density with the light-travel 
time across the emitting region. This allowed us to estimate the Lorentz factors
as well as viewing and opening angles for the jets of all of the sources in our sample. 
Here we apply these parameters in our interpetation of the multifrequency 
polarization properties of the sources. 
Because of difficulties in coordinating 
the schedules at the different telescopes, our observations at 
different wavelengths are contemporaneous (within two days to two weeks
of each other) rather than exactly simultaneous. Nevertheless, the program provides
the richest multi-epoch, multiwaveband polarization dataset compiled to date.

\section{Observations and Data Reduction}
We have measured the total flux densities and polarization 
of the objects in our sample in four spectral regions: at 7~mm (43~GHz),
3~mm (86~GHz), 0.85/1.3~mm (350/230~GHz), and optical wavelengths 
(an effective wavelength of $\sim$600--700~nm). 
 
\subsection{Optical Polarization Observations}
We carried out optical polarization and photometric measurements at several 
epochs using the Two-Holer Polarimeter/Photometer \citep{SSS85} with the Steward
Observatory 1.5~m telescope located on Mt. Lemmon, Arizona, and 1.55~m telescope
on Mt. Bigelow, Arizona. This instrument uses a
semi-achromatic half-waveplate spinning at 20.65~Hz to modulate incident
polarization, and a Wollaston prism to direct orthogonally polarized beams
to two RCA C31034 GaAs photomultiplier tubes. For all polarization measurements
we used either a 4\farcs 3 or 8\farcs 0 circular aperture and, except
for 3C~273, no filter.  The unfiltered observations sample the
polarization between $\sim$320 and 890~nm, with an effective central wavelength
of $\sim$600-700 nm, depending on the spectral shape of the observed object.
In the case of 3C~273 a Kron-Cousins $R$ filter
($\lambda_{eff} \sim 640$~nm ) was employed to
avoid major unpolarized emission-line features \citep{SSA93}.  
We accomplished sky subtraction by
nodding the telescope to a nearby ($<$30''), blank patch of sky
several times during each observation.
Data reduction of the polarimetry followed the procedure described in
\citet{SMITH92}.

When conditions were photometric, we obtained differential $V$-band photometry
($R$-band photometry in the case of 3C~273) of several objects,
employing either an 8\farcs 0 or 16\farcs 0 circular aperture.
We used comparison stars in the fields of the AGN \citep{SMITH85, SMITH98}
to calibrate nearly all of the photometry.
We determined the $V$ magnitudes of 3C~111 and 3C~120 on 1999 February 13
from the photometric solution provided by observations
of equatorial standard stars \citep{LAND92}.

Table \ref{TAB_Opt} lists the results of the optical observations. The columns correspond
to (1) source, (2) epoch of observation,
(3) filter bandpass of the photometry, 
(4) apparent magnitude, 
(5) flux density $I_{\rm opt}$, 
(6) filter bandpass of the polarimetry, where $W$ denotes 
unfiltered, or ``white-light'' measurements, 
(7) degree of linear polarization $m_{\rm opt}$, and (8) polarization position 
angle $\chi_{\rm opt}$.
We have corrected the degree of linear polarization for the statistical bias 
inherent in this positive-definite quantity by using the method of \citet{WK74}.
Usually this correction is negligible because of the high signal-to-noise (S/N)
ratio of most measurements.

Because of the high observed optical polarization levels and high galactic
latitudes of the majority of the sources, interstellar polarization (ISP) is not a
significant concern in most cases. Interstellar polarization from dust within
the Milky Way Galaxy does appear to be a major component of the observed
optical polarization for 3C~111 ($b$ = $-8.8^\circ$) and 
3C~120 ($b$ = $-27.4^\circ$).
A star $\sim 50''$ west of 3C~111 shows very 
high polarization ($m = 3.6\% \pm 0.2$\%; $\chi = 123^{\circ}\pm 2^{\circ}$);
this was used to correct the polarimetry for ISP in the line of sight to 
the radio galaxy. The corrected measurements suggest that the intrinsic
polarization of 3C~111 was typically $<$3\% throughout the monitoring campaign.  
The average polarization for five stars within 8 arcmin of 3C~120 yields 
$m = 1.22\% \pm 0.06$\% and $\chi = 98^{\circ} \pm 1^{\circ}$. We have used 
this ISP estimate to correct the observed polarization of 3C~120 for interstellar
polarization and, as for 3C~111, list the corrected measurements in Table
\ref{TAB_Opt}. The corrected values indicate that 3C~120 had very low 
polarization ($m_{\rm opt} < 0.5$\%) throughout the monitoring program. 

\citet{IMPEY89} have determined the interstellar polarization along the line
of sight to 3C~273 to be small, but the low levels of polarization
observed for this quasar require that this estimate of the ISP be subtracted 
from the $R$-band measurements. Table \ref{TAB_Opt} 
lists the corrected polarization for 3C~273.

\subsection{Submillimeter Polarization Observations}
We performed observations at 1.35 and 0.85~mm with the James Clerk Maxwell
Telescope (JCMT) located on Mauna Kea, Hawaii using the Submillimetre
Common User Bolometer Array \citep[SCUBA, ][]{HOL99}
and its polarimeter \citep{SCUBA03}. The initial plan was to observe exclusively
at 1.35~mm because the sources are almost always brighter, the atmospheric opacity 
is lower, and the sky is more stable than at 0.85~mm. However, failure of the
SCUBA filter drum in November 1999 forced a switch to 0.85~mm thereafter. 
The polarization properties of blazars tend to be very similar at millimeter 
and submillimeter wavelengths 
\citep{NAR98}, so the modest change in wavelength should not affect our 
analysis. For convenience we will refer to the data obtained with the JCMT
as ``1~mm'' data.

The SCUBA polarimeter consists of a rotating quartz half-waveplate and a
fixed analyzer. During an observation,
the waveplate is stepped through 16 positions, and photometric data are taken
at each position. One rotation takes $\sim$360~s to complete, and the
procedure results in a sinusoidally modulated signal from which the Stokes
parameters are extracted. A typical observation consists of 5--10
complete rotations of the waveplate. We achieved flux calibration in the
standard manner with observations of planets or JCMT secondary calibrators. We
measured the instrumental polarization ($\sim 1$\%) during each run by 
making observations of a compact planet, usually Uranus, which is assumed to 
be unpolarized at millimeter/submillimeter wavelengths.

We performed the initial (nod compensation, 
flat-field, extinction correction and sky noise removal, if appropriate) 
and polarimetric stages of the data reduction using the standard SCUBA data
reduction packages, $SURF$ and $SIT$ \footnote
{http://www.jach.hawaii.edu/JCMT/continuum/data/reduce.html}. The
Stokes parameters were extracted by fitting sinusoids to the data, either 
half-cycle (8 points) resulting in two estimates, or full-cycle (16 points) 
yielding only one estimate but generally giving better results with noisy 
data. We removed spurious measurements by performing a Kolmogorov-Smirnov 
test on the collated data and then calculated the degree and position angle of
the polarization, corrected for instrumental polarization and parallactic angle. 
Table \ref{TAB_JCMT} lists the data with the columns corresponding to 
(1) source, (2) epoch of observation, (3) wavelength $\lambda$, 
(4) flux density $I_{\rm 1mm}$, (5) degree of linear polarization $m_{\rm 1mm}$,
and (6) polarization position angle $\chi_{\rm 1mm}$. 
As with the optical measurements, the values of $m_{\rm 1mm}$
have been corrected for statistical bias.

\subsection{3-mm Polarization Observations}
From April 2000 to April 2001 we monitored the sources at 86 GHz using the 
linear polarization system on the the Berkeley-Illinois-Maryland Array (BIMA) 
in Hat Creek, California during unsubscribed telescope time. The data quality 
was quite variable, and typical total integration times were about 10 minutes 
per source. We omit from our analysis data taken during extremely high
atmospheric phase variations or bad opacity conditions. We observed planets (Mars,
Uranus, and Venus), H~II regions (W3OH and MWC349), and 3C~84 as
quality checks and for flux calibration.

We analyzed the data by mapping all four Stokes parameters for each
sideband separately, producing images of total intensity,
linearly polarized flux density, degree of polarization, and 
polarization position angle for each source at each epoch. Data
having S/N $<$3 for both the total and linearly polarized intensity
images at the central pixel of the source were omitted. 
We determined the total and polarized flux densities by fitting a Gaussian
brightness distribution to the corresponding image. 
Table \ref{TAB_BIMA} contains the results of the observations after correction
of the degree of polarization for statistical bias: 
(1) source, (2) epoch of observation,
(3) flux density $I_{\rm 3mm}$, (4) degree of linear polarization $m_{\rm 3mm}$,
and (5) polarization position angle $\chi_{\rm 3mm}$.

\subsection{7-mm VLBA Observations}
We performed total and polarized intensity imaging at 43~GHz
with the Very Long Baseline Array (VLBA) at 17 epochs 
from 25 March 1998 to 14 April 2001. We describe the observations and data
reduction in detail in J05, where the total and polarized intensity
images are presented. Table \ref{TAB_VLBA} gives the results of fitting
the VLBI core seen in the total and polarized intensity images
by components with circular Gaussian brightness distributions (see J05). 
Columns of the table are as follows: (1) source, (2) epoch of observation,
(3) flux density in the core $I_{\rm 7mm}$,
(4) projected inner jet direction $\Theta_{\rm jet}$,
(5) degree of linear polarization in the core $m_{\rm 7mm}$, and (6)
polarization position angle in the core $\chi_{\rm 7mm}$.
We determine the projected inner jet direction from the position 
of the brightest jet component closest to, but at least one synthesized
beam width from, the core.

Table \ref{TAB_Pcomp} gives the polarization of jet components
downstream of the VLBI core that are either (i) brighter than or comparable 
to the core at least at one epoch ($I^{\rm comp}\ge$0.5~$I_{\rm 7mm}$) or 
(ii) have detectable polarization at three or more epochs.  
These data allow us to probe the physics of the strongest disturbances that
propagate down the jet. Table \ref{TAB_Pcomp} contains: (1) source, 
(2) epoch of observation, (3) designation of the component, which follows that of J05,
(4) flux density in the component $I^{\rm comp}_{\rm 7mm}$, (5) distance from the 
core $R$, (6) position angle relative to the core $\Theta$, (7) 
degree of linear polarization $m^{\rm comp}_{\rm 7mm}$, and 
polarization position angle $\chi^{\rm comp}_{\rm 7mm}$, in the component. 
Values of the degree of polarization in both tables are corrected for 
statistical bias.

\section{Observed Characteristics of Polarization}
\begin{figure}
\epsscale{1.0}
\plotone{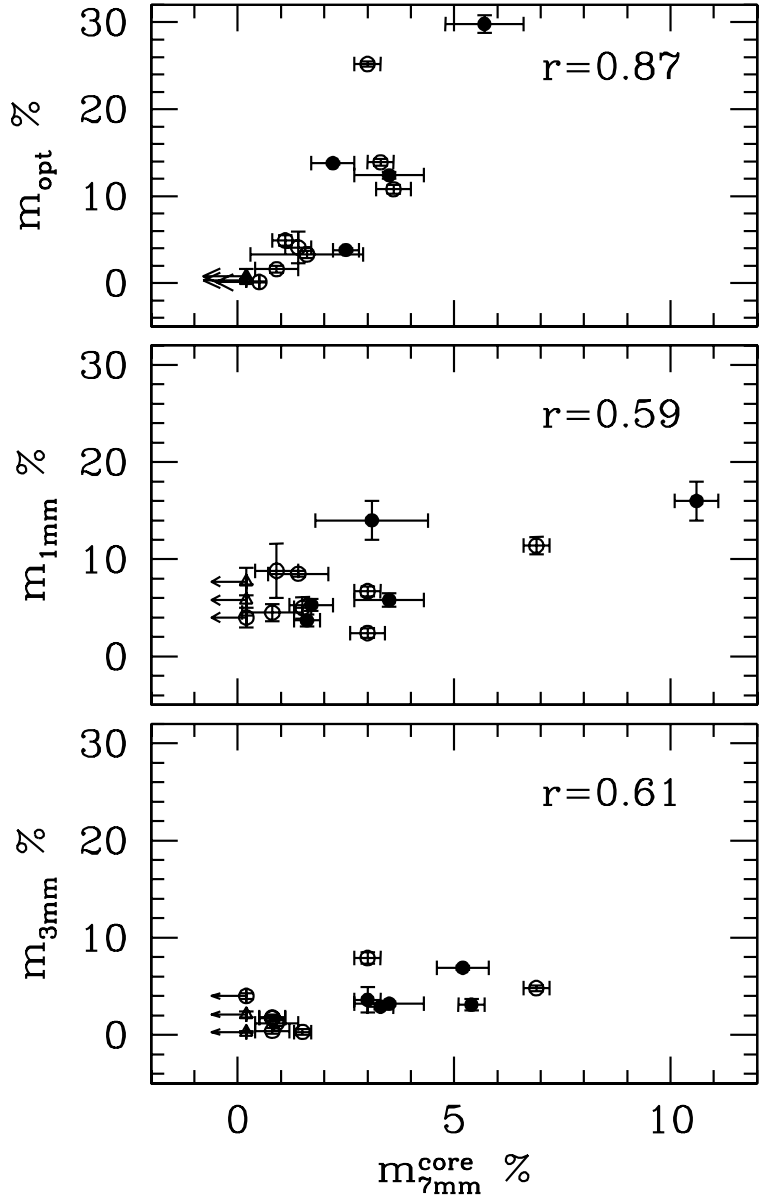}
\caption{Degree of polarization at optical, 1~mm, and 3~mm wavelengths 
from the whole source vs.
degree of polarization measured in the VLBI core at 7~mm for the most 
nearly simultaneous pair of observations for each source. The linear coefficient 
of correlation, $r$, is given in each panel.
Symbols denote the quasars (open
circles), BL~Lac objects (filled circles), and radio galaxies (triangles).} 
\label{P_sim}
\end{figure}

Tables \ref{TAB_Opt}-\ref{TAB_VLBA} show that the linear polarization for
all objects at most wavelengths varies significantly both in degree and position
angle. 
Figure \ref{P_sim} represents the relationship between the degree of polarization
at optical, 1~mm, and 3~mm wavelengths from the whole source with respect to the
degree of polarization measured in the VLBI core at 7~mm. The data plotted
in Figure \ref{P_sim} correspond to the closest observations obtained at opt-7~mm,
1-7~mm, and 3-7~mm wavelengths for each source. These measurements
are simultaneous within 1-3 days.
There is a statistically significant correlation at a significance level 
$\epsilon$=0.05 \footnote{Throughout the paper
we calculate coefficients of linear correlation and use the method of 
\citealt{BL72} for testing the significance of these coefficients. 
The hypothesis that 
there is no correlation between two variables, $r$=0, can be rejected
at a significance level $\epsilon$ 
if $t=|r/\sqrt{1-r^2}|\sqrt{N-2}\ge t_{\epsilon/2;N-2},$
where $t_{\epsilon/2;N-2}$ is the percentage point of the $t$-distribution
for $N-2$ degrees of freedom and $N$ is the number of observations.} between fractional polarization 
in the VLBI core and degree of polarization at short wavelengths.
The optical polarization maintains the strongest connection to the polarization
in the VLBI core ($r$=0.87, where $r$ is the linear coefficient of correlation).   
This result confirms the strong correlation between the polarization
level of the radio core at 7~mm and overall optical polarization
found by \citet{LS00}. Their sample included quasars with both high and 
low optical polarization.
However, it raises the question of why the optical polarization shows
a better connection to the polarization in the VLBI core than does
the polarization at 1~mm.    

\subsection{Polarization Variability}
We have computed the fractional polarization variability index 
$V^{\rm p}_\lambda$ at each frequency using the definition employed by 
\citet{AAH03a}:
\begin{equation}
V^p_{\lambda}=\frac{(m^{\rm max}_{\lambda}-\sigma_{m^{\rm max}_{\lambda}})-(m^{\rm min}
_{\lambda}+\sigma_{m^{\rm min}_{\lambda}})}{(m^{\rm max}_{\lambda}-\sigma_{m^{\rm max}_{\lambda}})+(m^{\rm min}_{\lambda}+\sigma_{m^{\rm min}_{\lambda}})}, \label{e1}
\end{equation}
where $m_\lambda^{\rm max}$ and $m_\lambda^{\rm min}$ are, respectively, the maximum and minimum
fractional polarization measured over all epochs at wavelength $\lambda$,
and $\sigma_{m^{\rm max}}$ and
$\sigma_{m^{\rm min}}$ are the corresponding uncertainties.  
This approach is especially justified in our case, since all objects were observed
over the same period of time, although the time intervals are different at 
different wavelengths ($\sim$2~yr in the optical, $\sim$1~yr at 3~mm, and $\sim$3~yr at 1~mm and 7~mm). 
Note that eq.~(\ref{e1}) can produce a negative variability index if the degree of
polarization or maximum change in polarization do not exceed the uncertainties in the
measurements. In this case the data fail to reveal polarization variability
and we set $V^{\rm p}=$0.
We introduce the polarization position angle variability index, $V^{\rm a}_\lambda$:
\begin{equation}
V^{\rm a}_\lambda = \frac{|\Delta{\chi_\lambda}|-\sqrt{\sigma_{\chi_\lambda^1}^2+\sigma_{\chi_\lambda^2}^2}}{90},\label{e2}
\end{equation}
where $\Delta{\chi_\lambda}$ is the observed range of polarization
direction and $\sigma_{\chi_\lambda^1}$, $\sigma_{\chi_\lambda^2}$ are
the uncertainties in the two values of EVPAs that define the range. 
We treat the $180^\circ$ ambiguities in EVPAs such that
$\Delta{\chi_\lambda}$ can not exceed $90^\circ$. As in the case of $V^{\rm p}$,
if $V^{\rm a}\le$~0 then we set $V^{\rm a}=$~0 and conclude that the observations were 
unable to measure variability in the polarization position angle. 
 
\begin{figure}
\epsscale{1.0}
\plotone{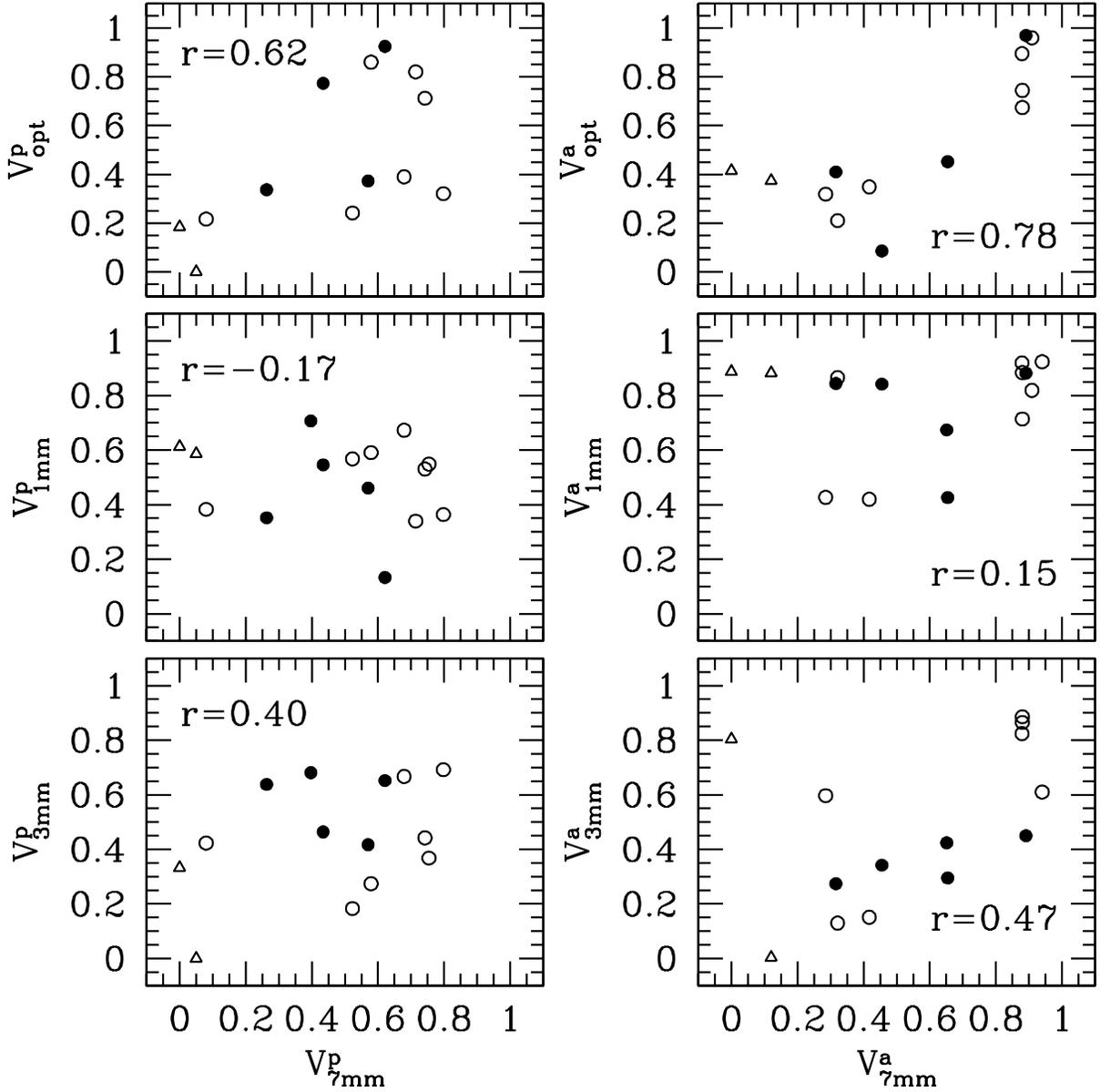}
\caption{{\it Left panel:} Dependence between polarization variability index 
at optical, 1~mm, and 3~mm wavelengths and polarization variability index
in the VLBI core at 7~mm. {\it Right panel:} Dependence between polarization 
position angle variability index at optical, 1~mm, and 3~mm wavelengths and 
polarization position angle variability index in the VLBI core at 7~mm.
The symbols are the same as in Fig. \ref{P_sim}.} \label{VpVa_index}
\end{figure}

Figure \ref{VpVa_index} shows the relationship between the polarization 
and position angle variability indices in the VLBI core and at short wavelengths. 
A statistically significant correlation occurs both between 
$V^{\rm p}_{\rm opt}$ and $V^{\rm p}_{\rm 7mm}$ and between $V^{\rm a}_{\rm opt}$ and 
$V^{\rm a}_{\rm 7mm}$ ($r$=0.62 and $r$=0.78, respectively). 
This result indicates that changes in the ordering of
magnetic fields in the optical region and VLBI core are correlated, which
suggests that the polarization variability has the same origin at the two wavebands.
The correlation between the values at 3~mm and in the VLBI core 
might be affected by the region of the jet lying somewhat outside the VLBI core 
that contributes significantly to the 3~mm polarized emission. There is
no correlation between variability indices at 1~mm and in the VLBI core.

\subsection{A Polarization-Based Classification Scheme}
We use the polarization variability indices to classify the sources
in our sample with respect to their polarization properties.
This classification scheme differs somewhat from the classical 
separation of AGN into radio galaxies, quasars, and BL~Lac objects, 
which we have used to discuss kinematics in the parsec-scale jets (J05). 
The new categorization reveals significant
differences in the polarization properties (see \S 7) among the 
introduced groups that would be diluted in the traditional scheme.

\begin{figure}
\epsscale{1.0}
\plotone{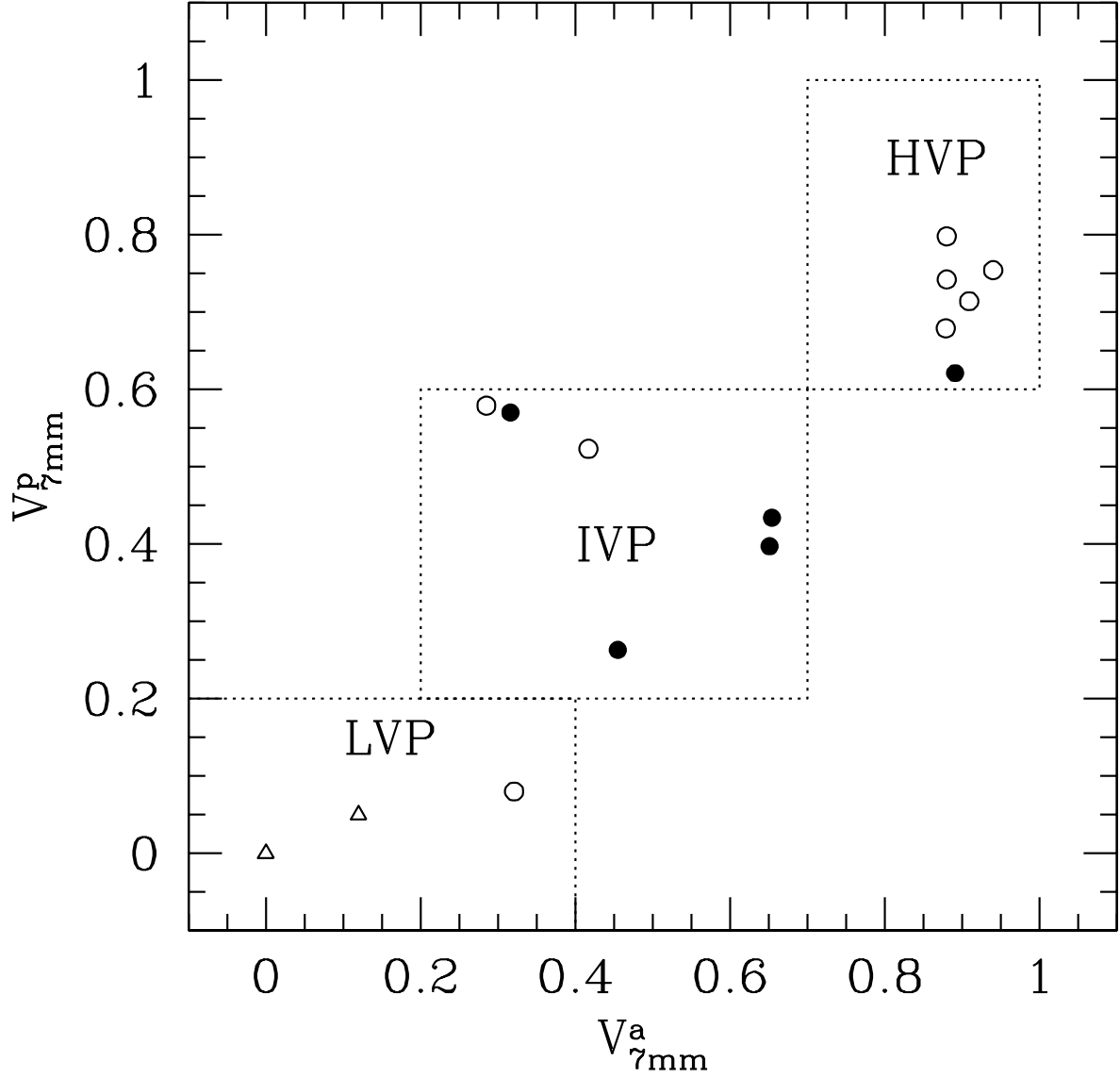}
\caption{ Connection between polarization and position angle variability indices
in the VLBI core at 7~mm. Symbols denote the quasars (open
circles), BL~Lac objects (filled circles), and radio galaxies (triangles)}. \label{VV7}
\end{figure}

Figure \ref{VV7} shows that the fractional polarization and polarization 
position angle variability indices in the 7~mm core are strongly correlated ($r$=0.83) 
in a way that produces a clear separation of the sources into three groups:
LVP --  low variability of polarization in the radio core, 
$V^{\rm p}_{\rm 7mm}\le$0.2, $V^{\rm a}_{\rm 7mm}\le$0.4; 
IVP -- intermediate variability of polarization,
$0.2<V^{\rm p}_{\rm 7mm}\le$0.6, $0.2<V^{\rm a}_{\rm 7mm}\le$0.7; 
and HVP -- high variability of polarization in the radio core, 
$V^{\rm p}_{\rm 7mm}>$0.6, $V^{\rm a}_{\rm 7mm}>$0.7. (We note, however, that 
low electric vector variability indices in the LVP group could result 
from large uncertainties in EVPAs connected with low fractional polarization.)

The LVP group includes the two radio galaxies 3C~111 and 3C~120 plus 
the low optically polarized quasar 3C~273. All three sources display
low fractional polarization in the optical band and in the 7~mm core. 
However, in the quasar 3C~273 the optical synchrotron emission is known to be 
variable and to have $m >$10\% if dilution from essentially unpolarized 
non-synchrotron components such as the big blue bump is taken into account 
\citep{IMPEY89}. All three LVP sources are highly polarized and variable 
at 1~mm, and the fractional
polarization of 3C~273 at 3~mm might be as high as 4\%.

The IVP group consists of four (out of five) BL~Lac objects and 
two highly optically polarized quasars, 3C~279 and 3C~345. 
This group, therefore, includes extremely highly polarized blazars, whose 
polarization in the optical and at 1~mm can exceed 30\% and whose
polarization at 3~mm and in the VLBI core does not drop below 2-3\%. Note
that \citet{NAR98} have found no differences in polarization properties
of BL~Lac objects and compact flat-spectrum quasars at short mm and sub-mm
wavelengths.

The majority of the quasars (five out of eight) and one BL~Lac object, 
OJ~287, form the HVP group. The linear polarization of the VLBI 
core in these sources can be very low---at the noise level---or 
as high as $\ge$5\%. 
The polarization variability indices, $V^{\rm p}$ at 3~mm, 1~mm, and  
optical wavelengths are similar to those for the IVP group, although $V^{\rm a}$ 
indices indicate more dramatic polarization position angle variability
at all wavelengths. We did not observe high
polarization at short mm wavelengths ($m_{\rm 3mm}\ge$10\% and 
$m_{\rm 1mm}\ge$15\%) as in the IVP group, although very high optical polarization 
($m_{\rm opt}\ge$20\%) might occur in these sources as often
as in IVP sources \citep{HT80}. 

\begin{figure}
\epsscale{1.0}
\plotone{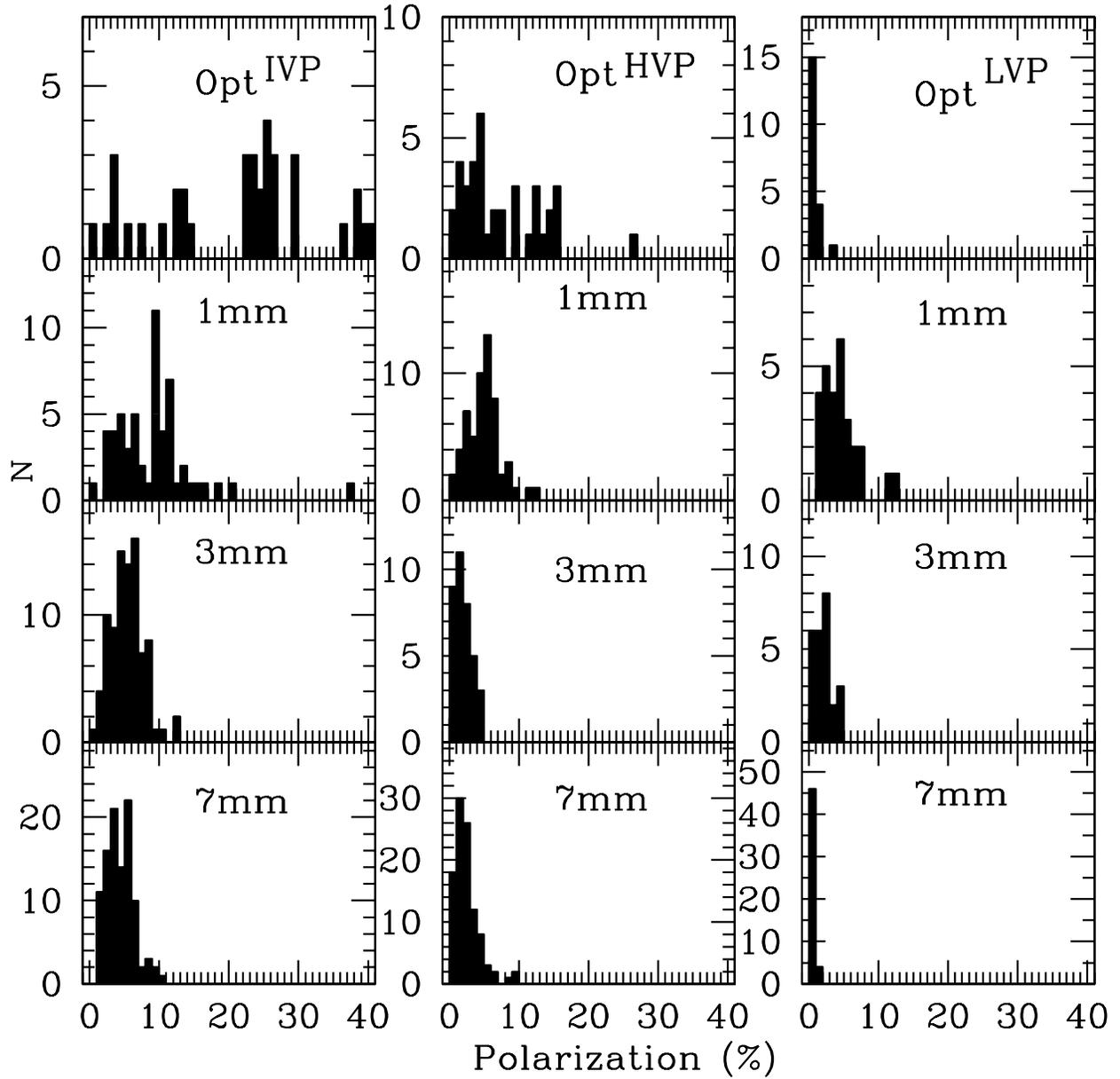}
\caption{ Distributions of degree of linear polarization in
the IVP ({\it left}), HVP ({\it middle}), and LVP ({\it right})
variability groups at optical, 1~mm, and 3~mm wavelengths and in the
VLBI core at 7~mm.}   \label{h_POL}
\end{figure}

Figure \ref{h_POL} presents the distributions of fractional polarization for all 
of our measurements, separated according to the variability groups. 
Comparison of the distributions shows that (i) the peak of the
distribution in the IVP group occurs at a higher percentage polarization
than for the other two groups, independent of wavelength; and (ii) for a given variability group,
the peak of the distribution shifts to higher fractional polarization at 
shorter wavelengths except for the LVP sources,
where the optical polarization is similar to the polarization in the VLBI core.

\section{Total and Polarized Flux Density Spectra}
We have constructed single-epoch total and polarized flux density spectra based on
measurements at different wavelengths obtained within 2 weeks of each other
(except 1803+784 and 1823+568, where, at best, there is nearly a month 
between observations at different wavelengths). 
We have corrected the optical flux densities shown in Table \ref{TAB_Opt} 
for galactic extinction using values of $A_\lambda$ compiled by the 
NASA Extragalactic Database. The total flux densities at 7~mm are corrected 
for possible missing flux density in the VLBA images using single-dish
observations as described in J05. 
The measurements at 7~mm include the core and components within
the 1\% contours of the peak total intensity at a given epoch (J05),
$I_{\rm jet}=I^{\rm core}+\sum_i{I^{\rm comp}_i}$.
The polarized flux density at 7~mm is integrated over the same VLBA image.
Therefore, $I^p_{\rm jet}=\sqrt{Q_{\rm jet}^2+U_{\rm jet}^2},
Q_{\rm jet}=Q^{\rm core}+\sum_i{Q^{\rm comp}_i},
U_{\rm jet}=U^{\rm core}+\sum_i{U^{\rm comp}_i},$ 
$m_{\rm jet}=I^p_{\rm jet}/I_{\rm jet},$ and
$\chi_{\rm jet}=0.5~\tan^{-1}(U_{\rm jet}/Q_{\rm jet})$, where $I^{\rm core}$,
$Q^{\rm core}$, and $U^{\rm core}$ are the Stokes parameters of the core,  
and $I^{\rm comp}_i$, $Q^{\rm comp}_i$, and $U^{\rm comp}_i$ are the Stokes parameters of 
a given polarized jet component.
Figure \ref{Spectra} presents the total and polarized spectral energy distributions
(SED), while Table \ref{TAB_Spec} lists the spectral indices $\alpha_{\rm mm}$ 
(based on total flux densities at
the three mm wavelengths) and $\alpha_{\rm opt/1mm}$,
where $S_\nu\propto \nu^{-\alpha}$. We calculate the spectral indices 
$\alpha^{\rm p}_{\rm mm}$ and  $\alpha^{\rm p}_{\rm opt/1mm}$ using the 
corresponding polarized flux densities, $S^{\rm p}_\nu\propto \nu^{-\alpha^p}$.

\begin{figure}
\epsscale{1.0}
\plotone{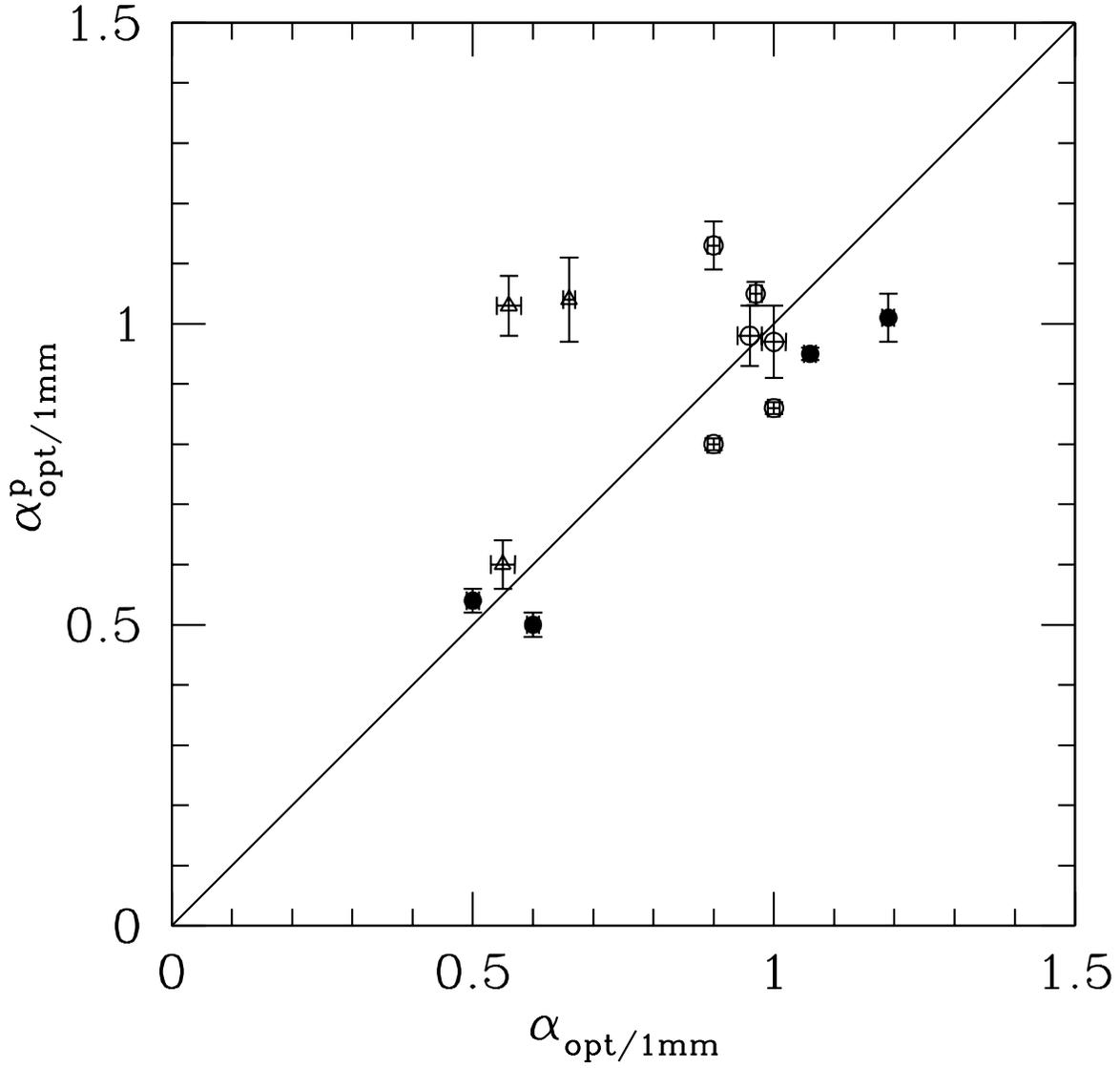}
\caption{Dependence between polarized and total flux spectral indices
calculated between the optical and 1~mm wavelengths
(open circles - HVP sources, filled circles - IVP sources, triangles - LVP sources).
} \label{Alpha}
\end{figure}

Table \ref{TAB_Spec} shows that the total flux density spectra at mm-wavelengths
are flat, independent of the type of source, with the majority of spectral indices 
having $|\alpha_{\rm mm}|<$0.5. This confirms that the mm-wave emission is partially 
optically thick. The optical-mm spectra of the HVP sources are steep,
with $\alpha_{\rm opt/1mm}\sim$1, while the LVP sources possess much flatter
optical to 1~mm spectra. In the IVP group $\alpha_{\rm opt/1mm}$ depends on whether
the source is a quasar ($\alpha_{\rm opt/1mm}\sim$1) or BL~Lac object
($\alpha_{\rm opt/1mm}\sim$0.5). The polarized flux density spectra at mm
wavelengths are flat or inverted, which indicates 
an increase of polarized emission with frequency at mm wavelengths.
Figure \ref{Alpha} reveals a strong correlation
between $\alpha^{\rm p}_{\rm opt/1mm}$ and $\alpha_{\rm opt/1mm}$ ($r$=0.92). However, two sources from the LVP group (3C~120 and 3C~273)
deviate greatly from the dependence $\alpha^{\rm p}_{\rm opt/1mm}=\alpha_{\rm opt/1mm}$
(solid line in Fig. \ref{Alpha}),
with a significantly steeper polarized flux density spectral index than for
the total flux density. This suggests that at least two emission components are
present in the optical region, one of which is unpolarized. 
\citet{IMPEY89} find an increase in the degree 
of polarization and higher variability of the Stokes parameters in 3C~273
at longer optical wavelengths. 
Such behavior is expected if the optical emission consists of a variable synchrotron
component plus a relatively static blue unpolarized continuum source, such as the 
big blue bump. The decomposition of the optical synchrotron spectrum 
from other optical emission sources is readily seen in the 
spectropolarimetry of 3C~273 \citep{SSA93}. In the case
of 3C~120, the strong dilution of the optical synchrotron polarization 
is likely caused by host galaxy starlight included within the measurement 
aperture. 
 
In the majority of the HVP and IVP sources, the spectral indices
$\alpha^{\rm p}_{\rm opt/1mm}$ and $\alpha_{\rm opt/1mm}$ are similar 
to each other. This suggests that a single 
synchrotron component is responsible for the total flux and polarized
continuum from millimeter to optical wavelengths. 
However, three of the HVP sources (0420$-$014, 1510$-$089, and 3C~454.3) 
have a slightly steeper polarized flux density spectrum relative to the 
total flux density spectrum 
(this was found previously as well by \citealt{SMITH88}). 
In the IVP sources the opposite trend prevails,
except for BL~Lac. A steeper polarized spectrum in the quasars indicates the 
possible presence of a non-synchrotron component, such as that 
observed for 3C~273, although its contribution to the total flux should 
be $\lesssim$10\% relative to the synchrotron emission. 
A flatter polarized spectrum likely rules out any significant contribution
by a non-synchrotron component, although two (or more)
synchrotron components might coexist in the IVP sources. For BL~Lac,
contribution to the optical flux from the host galaxy might help to explain 
thr slightly steeper $\alpha^{\rm p}_{\rm opt/1mm}$ than the total flux 
spectral index.

\section{Faraday Rotation in the Inner Jet}
Figure \ref{Spectra} presents measurements of the
polarization angle $\chi$ at different frequencies obtained within
2~weeks of each other and shows that 
the direction of mm-wave polarization rotates with wavelength.
In the HVP and IVP sources the EVPA at 7~mm displayed in the figure 
corresponds to that of the VLBI core only, while in the LVP sources it
corresponds to $\chi_{\rm jet}$, as described in \S 4. 
We attribute this rotation to Faraday rotation 
by a foreground screen close to the VLBI core \citep{ZT04,AWH05}. 
We define the rotation measure $RM$ by assuming that at 1~mm 
we see the unrotated direction of the polarization and that the emission 
from 1~mm to 7~mm propagates through the same screen.

\subsection{Faraday Rotation in the HVP and IVP Sources}
\begin{figure}
\epsscale{1.2}
\vspace{-3.5cm}
\hspace{-3.5cm}
\plotone{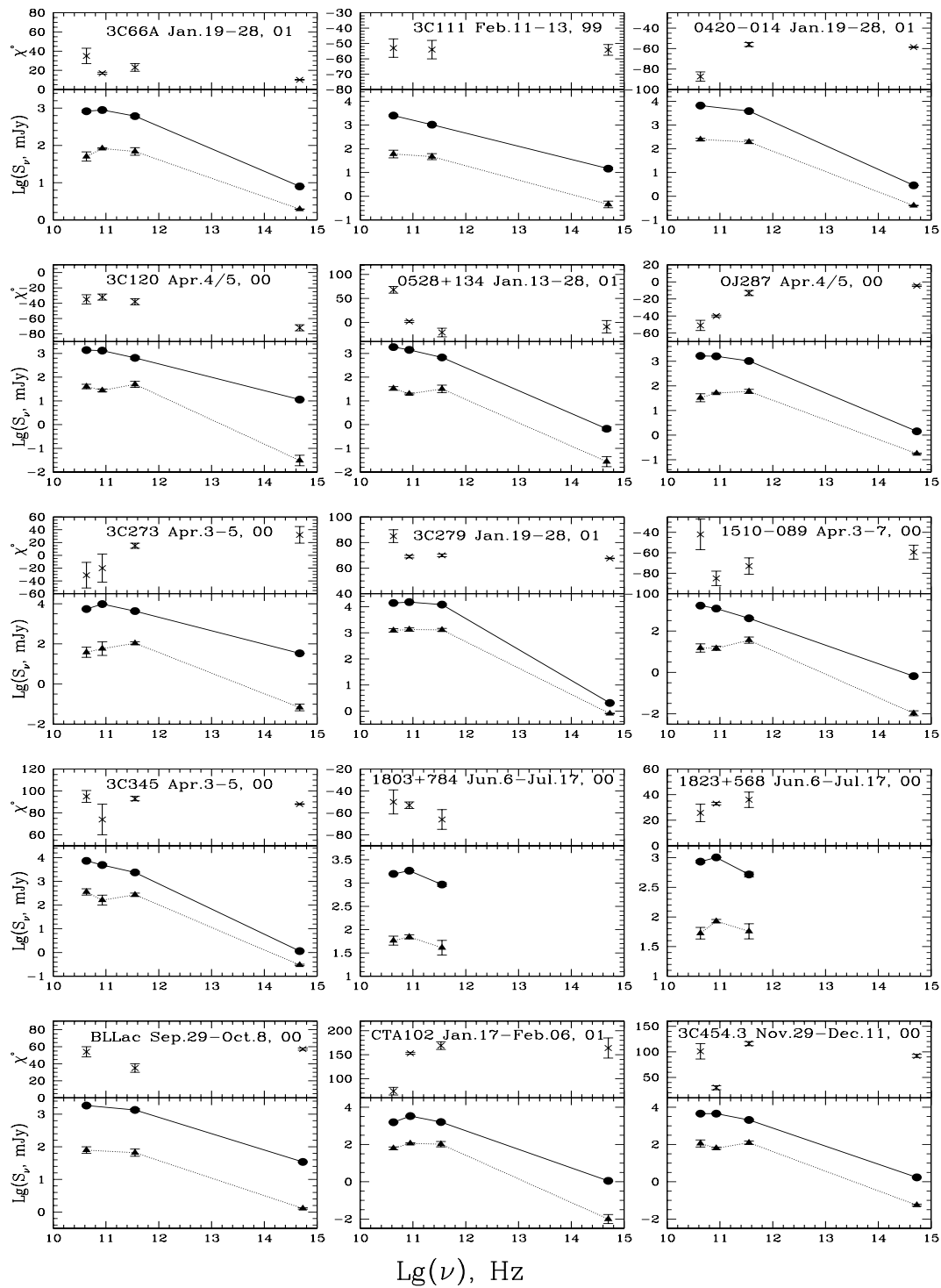}
\vspace{-4cm}
\caption{Total (filled circles) and polarized (filled 
triangles) flux density spectra and polarization position angle (crosses) measurements 
from optical to 7~mm wavelengths.}
\label{Spectra}
\end{figure}

\begin{figure}
\epsscale{1.2}
\vspace{-3cm}
\hspace{-3.5cm}
\plotone{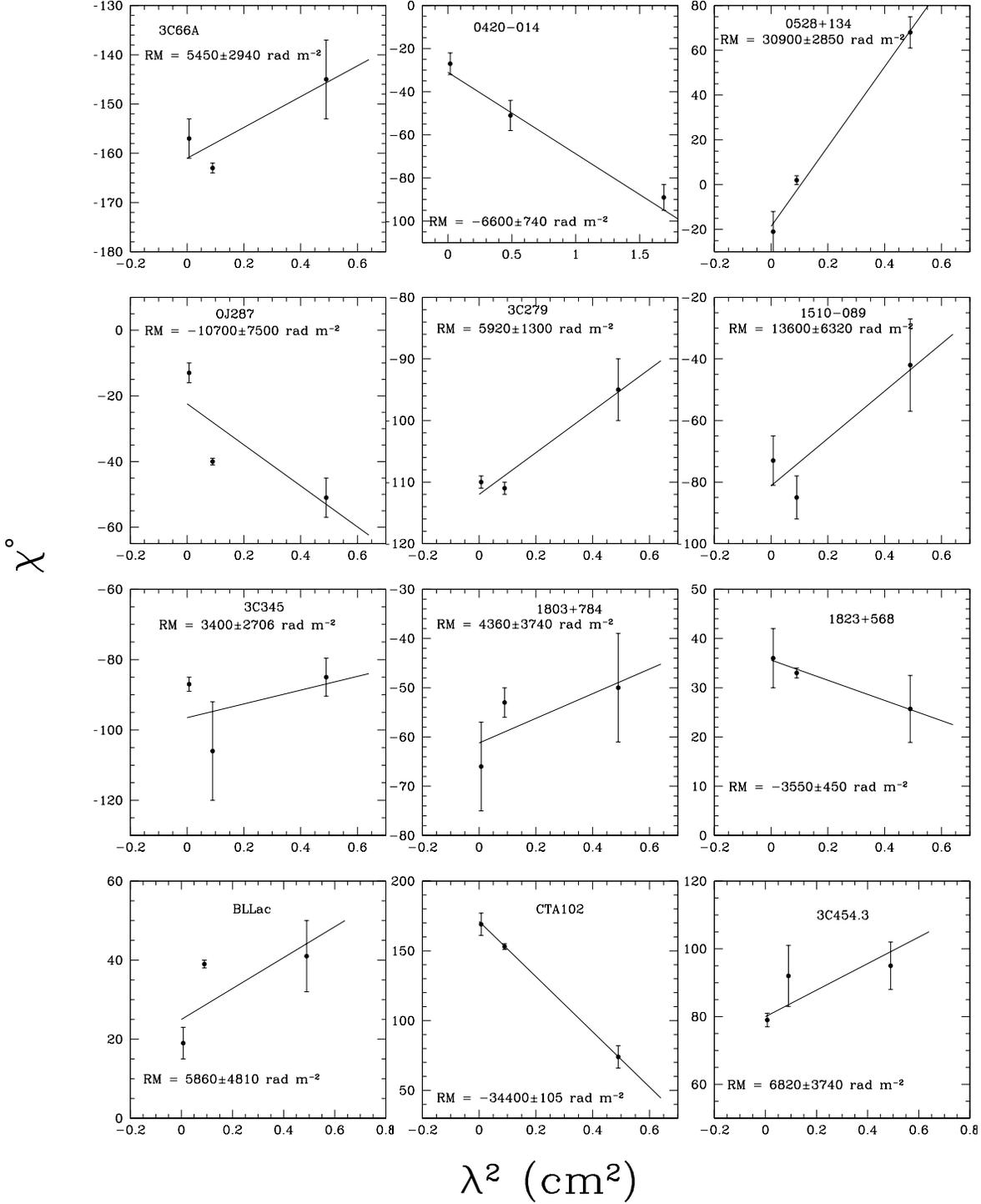}
\vspace{-4.5cm}
\caption{Dependence of polarization position angle on square of wavelength
in the IVP and HVP sources.
Solid line represents approximation of the dependence by a $\lambda^2$
Faraday rotation law.} \label{FRM}
\end{figure}

Figure \ref{FRM} shows the result of fitting the dependence of 
the polarization position angle on wavelength by a Faraday rotation law: 
$\chi=\chi_{\rm 1mm}+RM\lambda^2$. For 0420$-$014
our program does not contain simultaneous observations
at three mm wavelengths necessary to estimate $RM$.
We have supplemented our data by observations obtained with the VLBA
at 7~mm and 1.3~cm in the program BG073 \citep[see~][]{G00},
for which 0420$-$014 and 3C~454.3 were used as calibrators.
For 0420$-$014 we use our polarization measurement at 1~mm
on 1998 December 13 and EVPAs in the 7~mm and 1.3~cm core
on 1998 December 3 from BG073. The EVPA at 7~mm is consistent
with $\chi_{\rm 7mm}$ at epoch 1998 December 11 from our program.
For 3C~454.3 and BL~Lac we use observations 
different from those plotted in Fig. \ref{Spectra}.
For 3C~454.3 we use observations at 1~mm 
and 3~mm from our program (2000 June 6 and 8, respectively)
and 7~mm VLBA observations on 2000 June 9 (BG073).  
These observations are closer to each other than the measurements 
presented in Figure \ref{Spectra} 
(2000, November 29 -- December 11) and less affected by the variability 
of bright component $B6$ seen on the images (J05). We
have included the contribution of the component in the polarization at 7~mm
because $B6$ is brighter and more highly polarized than the core, and located 
within 0.2~mas of the core. For BL~Lac we use observations 
carried out over the period 2001 January 17-28 in order to have simultaneous 
measurements at three mm wavelengths. 
 
Table \ref{TAB_RM} gives the values of correction for EVPAs at 3~mm and 7~mm
along with the intrinsic rotation measure for each object, $RM^\circ=
RM~(1+z)^2$. Some of the $RM$ values  have large uncertainties that might
be attributed to several problems which our observations possess. First, 
the observations 
are not completely simultaneous, whereas the polarization position angle
can sometimes have short timescales of variability even at 7~mm \citep{FRANI07}.
Second, the polarization position angles at 7~mm correspond to those
of the VLBI core only, while the EVPAs at 1~mm and at 3~mm are from the
whole source. Although we have tested that, at epochs used
for estimates of the rotation measure, jet components 
contribute little to the polarization of the core 
(except 3C~454.3), their contribution to the 
polarized flux at 1~mm and 3~mm is unknown. 
Third, the structure of the 
magnetic field in the 7~mm core and in the regions emitting at shorter 
mm wavelength,- especially at 1~mm,-
might be different. Therefore, the observed rotation could be
an intrinsic property of the magnetic field structure. 

\begin{figure}
\epsscale{1.0}
\plotone{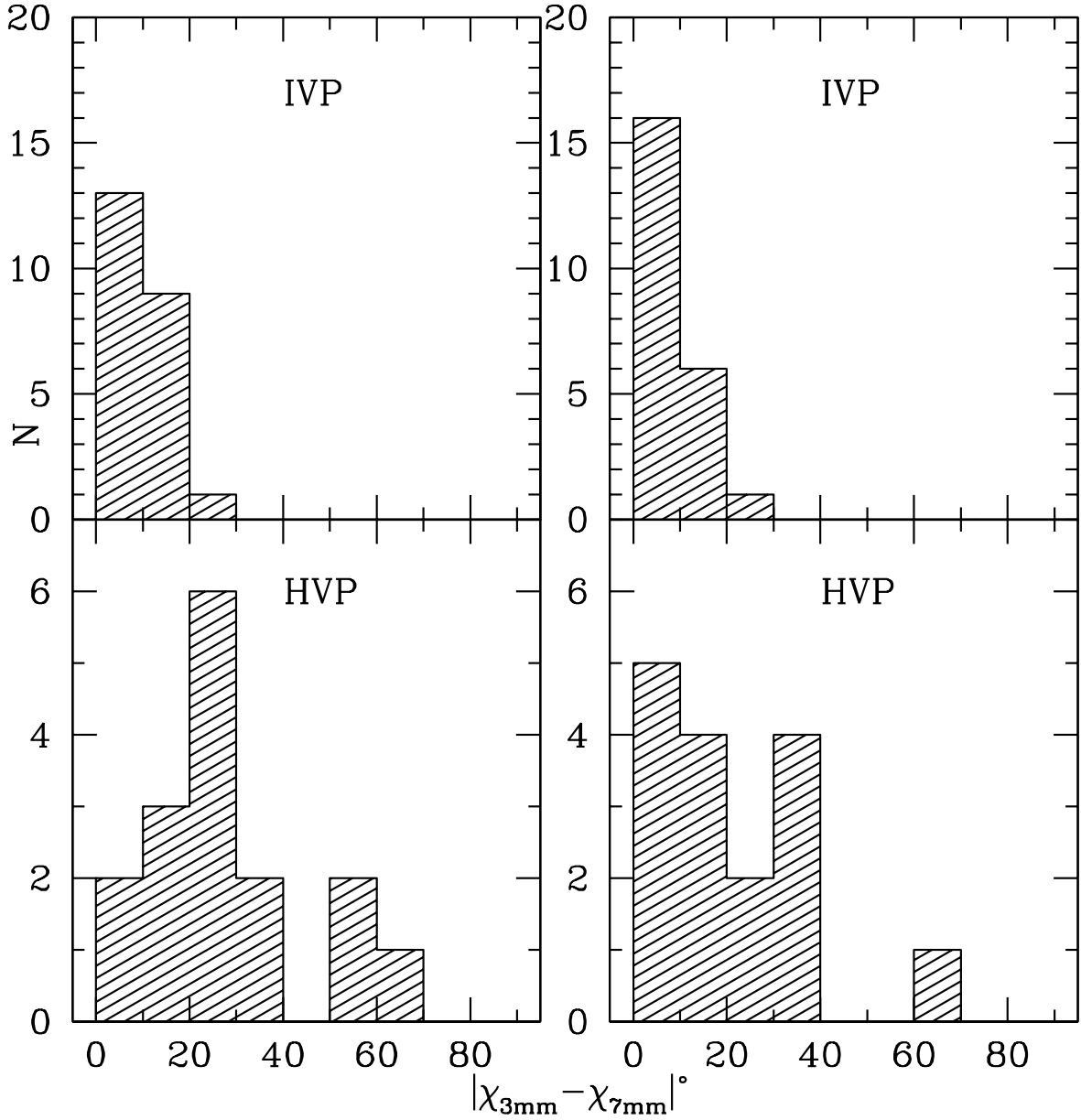}
\caption{Distributions of offsets between 
EVPAs at 3~mm and in the 7mm core before the $RM$ correction ({\it 
left panel}) and after the $RM$ correction ({\it 
right panel}). The distributions do not include the observations
used to calculate the $RM$.} \label{RM_dif}
\end{figure}

To test that the derived 
rotation measures are reliable, we construct the distributions of 
$\Delta\chi_{37}=|\chi_{\rm 3mm}-\chi_{\rm 7mm}|$
for all pairs of observations simultaneous within 2~weeks for $\chi_\lambda$
before and after $RM$ correction (Fig. \ref{RM_dif}).
Note that the distributions do not include the observations
used to calculate the rotation measures. This avoids the bias that these 
observations would introduce to the distributions. 
Ideally, the EVPAs after $RM$ correction should align
within 10$^\circ$-20$^\circ$, which corresponds to the uncertainties of our
measurements. This is a feature of both distributions (uncorrected
and corrected) for the IVP sources, reflecting the fact that in the 
IVP sources the uncertanties of EVPAs are comparable to the $RM$ 
correction values. However, EVPAs corrected for $RM$ exhibit a slightly 
better alignment than $\Delta\chi_{37}$ before the correction, as expected if the
EVPA rotation is caused by a Faraday screen.
In the HVP sources the $RM$ correction significantly improves 
alignment between $\chi_{\rm 3mm}$ and $\chi_{\rm 7mm}$:
69\% of uncorrected values of $\Delta\chi_{37}$ exceed 20$^\circ$
while only 44\% of corrected EVPAs have  $\Delta\chi_{37}> 20^\circ$.
The existence of a few large misalingments between EVPAs after the $RM$
correction can be attributed to variability of $RM$ in
the VLBI core of quasars \citep{ZT04}. This partially affects the
analysis performed in \S 7.

Table  \ref{TAB_RM} shows that the IVP sources have lower 
rotation measures, $\langle|RM^\circ|\rangle=(1.1\pm 0.3)\times 10^4$~rad~m$^{-2}$,
than the HVP sources, $\langle|RM^\circ|\rangle=(8.5\pm 6.5)\times 10^4$~rad~m$^{-2}$.
Two of the HVP blazars, 0528+128 and CTA~102, have EVPA corrections at
7~mm close to 90$^\circ$. A 90$^\circ$ flip in EVPA can be caused by the 
spectral properties of the source -- optically thin at 1~mm but
optically thick at 7~mm.  
However, according to \citet{PAH70} the degree of polarization should
decrease significantly (by a factor of $\sim$ 7) when the optical depth
to synchrotron self-absorption exceeds unity, while both quasars show only 
a moderate decrease of polarization from 1~mm to 7~mm at the epochs used 
to calculate $RM$ ($m_{\rm 1mm}=4.7\pm 1.6$\% and 
$m_{\rm 7mm}=1.8\pm 0.5$\% for 0528+134; $m_{\rm 1mm}=6.5\pm 1.9$\% and 
$m_{\rm 7mm}=4.0\pm 0.7$\% for CTA~102).

In a sample of 40 AGN \citet{ZT04} have determined rotation measures 
using VLBI images at 8-15~GHz that are lower than we derive at mm wavelengths.
The authors suggest that an external 
screen ``in close proximity to the jet'' is the most promising candidate for 
the source of the Faraday rotation. In this context, lower rotation measures obtained in 
the VLBI core at longer wavelengths might reflect a strong decrease 
in thickness of the screen with distance from 
the central engine. This is expected because the location of the core shifts with 
wavelength owing to optical depth effects. 
We assume a decreasing gradient in electron density of the screen, described by 
a power law: $N_e\propto~d^{-a}$, where $d$ is the distance from the central engine.
The rotation measure depends on the density and 
magnetic field parallel to the line of sight,  $B_\parallel$, 
integrated along the path through 
the screen to the observer: $RM\propto\int{N_e B_\parallel dl}$.
In the absence of a velocity gradient across the jet, the 
magnetic field along the jet scales as $B_{\rm z}\propto d^{-2}$
and the magnetic field transverse to the jet scales as $B_{\phi}\propto d^{-1}$
\citep{BBR84}. If the Faraday screen is at least mildly relativistic and has 
a helical field \citep{GAB06}, $B_{\phi}$ provides 
the main contribution to the magnetic field component along the line of sight,
therefore, $B_{\parallel}\propto d^{-1}$. 
This leads to $|RM|\propto d^{-a}$ under the approximation that $l\propto d$.
The location of the core from the central engine
depends on the frequency of the observation $\nu$ as derived, for example,  
in \citet{LOB98}:
$d_{\rm core,\nu} = (B_1^{k_b}F/\nu)^{1/k_r},$
where $B_1$ is the magnetic field at distance 1~pc from the
central engine, $F$ is a function of the redshift and parameters 
depending on the jet geometry and relativistic electron energy distribution,
and the equipartion value $k_r = 1$. For a source with a flat mm-wave spectrum
$d_{\rm core,\nu}\propto\nu^{-1}$, which yeilds the dependence of the 
rotation measure observed in the core on the frequency of observation  
as $|RM_{\rm core,\nu}|\propto\nu^a$.
Therefore, values of rotation measure obtained with different sets 
of frequencies should produce an estimate of the parameter $a$. 
We assume that the $RM_\circ$ derived from polarization observations of the core 
at frequencies $\nu_1$, $\nu_2$, and $\nu_3$, where $\nu_3$ is the lowest one,
yields the intrinsic rotation measure at the location $d_{\rm core,\nu_3}$. 
 
Table \ref{TAB_RM} contains the published rotation measures
obtained in the VLBI core at wavelengths longer than 7~mm, $RM^\circ_{\rm 15GHz}$
and $RM^\circ_{\rm 8GHz}$, and derived values of $a$.
For sources with $RM^\circ$ available at 
more than two wavelengths, $a$ is calculated using a least-squares method.
Table \ref{TAB_RM} does not show any difference in the values of $a$
between the HVP and IVP sources.  The average 
$\langle{a}\rangle=1.8\pm 0.5$ is similar to the value $a=2$ expected for outflow
in a spherical or conical wind. This implies that
an outflowing sheath wrapped around a conically expanding
jet is a reasonable model for the foreground screen, consistent
with the finding of \citet{ZT04}. The sheath can result from 
a mildly relativistic outer wind that emanates from the accretion disk 
and confines the inner relativistic jet \citep{GTB05}. According to
the $RM^\circ$ values in \ref{TAB_RM}, the thickness of the sheath is higher in 
the HVP sources with respect to the IVP sources, a situation that 
should assist in stronger confinement of the jet of HVP blazars. The latter 
may help to explain
the finding by J05 that the jet opening angles for quasars are smaller than 
for BL~Lac objects.
  
We have corrected the values of $\chi_{\rm 3mm}$ 
and $\chi_{\rm 7mm}$  in the HVP and IVP
sources for Faraday rotation by applying the corrections listed in 
Table \ref{TAB_RM} to the EVPAs given in Tables \ref{TAB_BIMA} and \ref{TAB_VLBA}.
The adjusted values of the EVPAs are used in the subsequent analysis.

\subsection{Faraday Rotation in the LVP Sources}
\begin{figure}
\epsscale{1.0}
\plotone{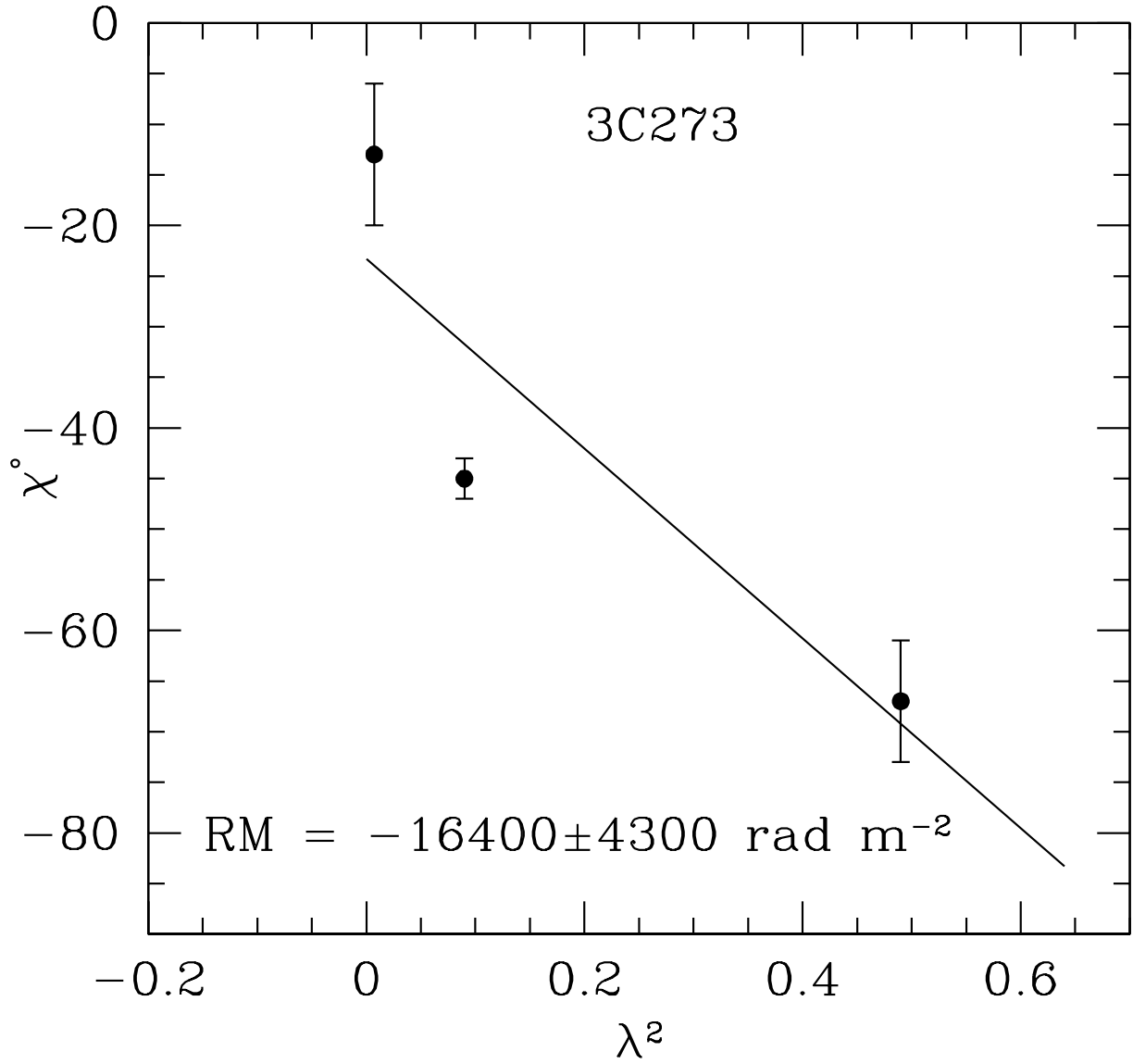}
\caption{Dependence of polarization position angle on square of wavelength
in 3C~273. Solid line represents approximation of the dependence by
a $\lambda^2$ Faraday rotation law.} \label{RM_LVP}
\end{figure}

In the LVP sources, the core at 7~mm is unpolarized, hence 
the polarization position angle of the inner jet is defined 
by the polarization of VLBI components within a few mas of the core. For the 
epochs shown in Figure \ref{Spectra} these are components $C1$ and $c1$
at $\sim$4~mas in 3C~111, components $t$ and $u$ at 1-2~mas in 3C~120, 
and component $B2$ at 1~mas in 3C~273. Figure \ref{RM_LVP} shows 
the $\lambda^2$ dependence of $\chi$ in 3C~273, which 
yields a high rotation measure, $RM=(1.6\pm 0.5)\times 10^4$~rad~m$^{-2}$.
The value is consistent with the high rotation measure obtained for component $Q6/W6$ 
from the 43/86~GHz images by \citet{AWH05}, $RM=2.1\times 10^4$~rad~m$^{-2}$.
However, a significant decrease in the density of the Faraday screen 
with distance from the core discussed above explains the good alignment
between the EVPA at 1~mm  and $\chi_{\rm jet}$ when components 
are seen farther downstream. This occurs for 3C~111 and 3C~120:
in 3C~111  $\chi_{\rm 1mm}=-54^\circ\pm 6^\circ$ and $\chi_{\rm jet}=-58^\circ\pm 10^\circ$;
in 3C~120 $\chi_{\rm 1mm}=-38^\circ\pm 4^\circ$ and $\chi_{\rm jet}=-35^\circ\pm 6^\circ$ . 
The implication is that the magnetic field in the 1~mm emission region have 
the same orientation as that in the jet features a few milliarcseconds from 
the core. Very low polarization in the VLBI core at 7~mm, the high 
rotation measure obtained close to the core in 3C~273, and consistency between the
direction of polarization in the inner jet and at 1~mm in 3C~111 and 3C~120, 
all suggest that in the LVP
sources the 1~mm and 3~mm cores have low polarization as well. The low
polarization of the core could be the result of either fine-scale turbulence
or depolarization by a very thick, inhomogeneous foreground screen 
with $RM>5\times$10$^5$~rad~m$^{-2}$, as suggested by \citet{AWH05}.

The EVPAs of the LVP sources are not corrected for $RM$ owing to
the complexities discussed above. We use simplification that, when strong 
polarized components are detected, they have left already the high 
$RM$ zone.
 
\section{Correlation Analysis} 
\subsection{Comparison of Polarization at Different Wavelengths} 
We have calculated linear coefficients of correlation $f^p$ between the polarized flux densities in the core at 7~mm
and overall polarized flux density measurements at (1) optical, 
$f^p_{o7}$, (2) 1~mm, $f^p_{17}$, and (3)  3~mm, $f^p_{37}$. These
apply to sources having data from at least three essentially 
simultaneous observations at two wavelengths.
(Here we consider the observation at two wavelengths ``simultaneous'' if they
are obtained within 2~weeks of each other.) 
We do the 
same for the correlation coefficients $r^m$ of the the degree 
of polarization in the core at 7~mm and the overall fractional polarization 
in each of the other wavebands, using the
same subscript designations. 
For some sources several measurements have been carried out
at high frequencies within two weeks of a few VLBA epochs. In these cases,
we use only the observation that is nearest to the corresponding VLBA epoch.
Table \ref{TAB_Pcor} contains coefficients of correlation $f^p_{o7}$, 
$f^p_{17}$, and
$f^p_{37}$ (columns 2-4) and $r^m_{o7}$, $r^m_{17}$, and $r^m_{37}$ (columns 5-7).
The number of points participating in the computation is indicated in parentheses.
The coefficients of correlation that are significant at a level
$\epsilon=0.1$ are indicated by bold font. 

The correlation coefficients between
optical polarization and polarization in the core at 7~mm  
correspond to neither positive 
nor inverse correlation despite otherwise similar overall polarization characteristics
(see \S 3). Blazars are known to 
have very short timescales of polarization variability at optical wavelengths
($<$1~day, \citealt{HT80,MSW87,MEAD90}). Because of this, we suspect that the 
lack of exact simultaneity and the small number of observations affect 
the correlation analysis in the optical band. 

The statistically significant coefficients 
of correlation listed in Table \ref{TAB_Pcor} indicate a strong 
positive correlation either 
between the polarized flux in the core at 7~mm and overall source at 1~mm and 3~mm, 
or between the fractional polarization in the core and at shorter mm wavelengths.  
However, the number of sources with such behavior corresponds to 25\% and 45\% 
for the 1-7~mm and 3-7~mm data, respectively. The absence of
statistically meaningful correlations for other sources might be explained by
either small-number statistics, e.g., for 0420$-$014 or 1510$-$089, 
or by contributions from jet components different from the core to 
the polarized flux at 1~mm and 3~mm.  
Table \ref{TAB_PCcor} gives the coefficients of correlation of the sources
for which inclusion of the jet components (see \S 4) changes 
the significance of the correlation. Table \ref{TAB_PCcor} shows
that including highly polarized components within a few milliarcseconds 
from the core dramatically improves the correlation in 1) all the LVP sources, 
2) two of the HVP sources (CTA~102 and 3C~454.3), and
3) none of the IVP sources. For 3C~279 the correlation 
coefficient between the 7~mm and 1~mm polarization increases ($r^m_{17}=0.51$), 
but does not change its significance substantially.

\subsection{Polarization - Total Flux Connection at Different Wavelengths}
We have computed coefficients of correlation between the degree of polarization and
total flux density, $r_\lambda$ (listed in Table \ref{TAB_PTcor}), and 
between the polarized and total flux densities, $f_\lambda$ (given in
Table \ref{TAB_PPTcor}). The coefficients are calculated for each source 
that has three or more simultaneous measurements of the 
two quantities at wavelength $\lambda$ (in this case the observations
are completely simultaneous).
Since the JCMT observations were performed at 1.35~mm and 0.85~mm, we have adjusted
flux densities and fractional polarization at 0.85~mm to 1.35~mm using the spectral
indices $\alpha_{\rm mm}$ and $\alpha^p_{\rm mm}$ provided in Table \ref{TAB_Spec}.
The data at 1~mm are supplemented by the measurements obtained by \citet{NAR98}
at 1.1~mm for sources common to both samples: 0420$-$014, 0528+134, OJ~287,
3C~273, 3C~279, 3C~345, 1823+568, BL~Lac, and 3C~454.3.
Coefficients of correlation $r_{\rm 7mm}$ and $f_{\rm 7mm}$ correspond 
to the VLBI core only. 
 
Table \ref{TAB_PTcor} reveals that, in general, the total flux density and 
percentage polarization are not well correlated. However, at 1~mm the one 
statistically significant coefficient (for BL~Lac) indicates a decrease 
of fractional
polarization when the source brightens, while at optical, 3~mm, 
and 7~mm wavelengths the few correlation coefficients that are meaningful
correspond to a positive 
correlation between the total flux density and degree of polarization.
An increase of the total flux at mm wavelengths is usually connected
with the emergence a new superluminal component from the VLBI core \citep{SAV02}.
The component might have a different direction of polarization than
the core and partially cancel the observed polarization, thus leading to a poor 
correlation between the total flux and degree of polarization.  
Table \ref{TAB_PPTcor}, which lists correlation coefficients between the total 
and polarized flux densities, yields a qualitatively different result. 
Significant coefficients of correlation indicate that the polarized flux density 
increases as the total flux density rises at all wavelengths. 
The correlation occurs for 44\%, 47\%, 46\%, and 77\% of sources at optical, 
1~mm, 3~mm, and 7~mm wavelengths, respectively. 

\begin{figure}
\epsscale{1.2}
\vspace{-3cm}
\hspace{-3.5cm}
\plotone{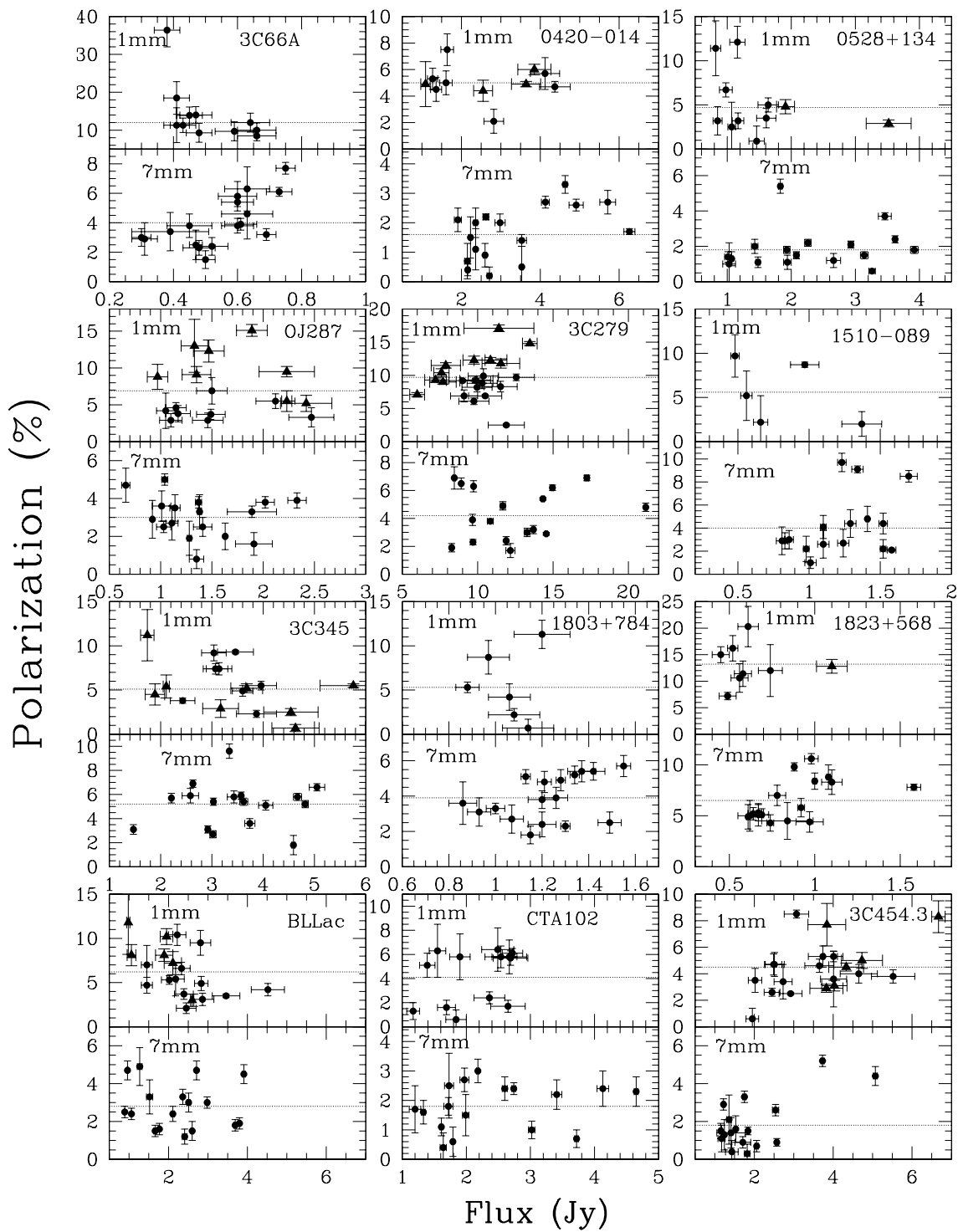}
\vspace{-4.5cm}
\caption{Fractional polarization vs. total flux density at 1~mm from the whole
source and in the core at 7~mm. Filled circles correspond to our observations and  
triangles repesent the data at 1.1~mm from \citet{NAR98}.}
\label{FPOL}
\end{figure}

Figure \ref{FPOL} shows the dependence between the percentage polarization
and total flux density at 1~mm and in the core at 7~mm for HVP and IVP sources.
For all 12 sources, the fractional polarization at maximum flux density
in the VLBI core exceeds the average degree of polarization (dotted line)
while at 1~mm eight sources show the inverse relation: the fractional polarization
at maximum flux density is lower than the average. The exceptions are
3C~279, CTA~102, and 3C~454.3, with all having jet components different
from the core that appear to contribute to the emission at 1~mm.
1803+784 is another exception - it has the smallest range of flux 
variability at 1~mm.

Despite poor correlation between the total flux density 
and percent polarization, - which partially can be affected by complex 
polarization structure of the mm-wave core region during ejection of 
a new VLBI component - the data support the inference that the
fractional polarization at 1~mm tends to decrease when a source brightens. 
In contrast, in the core at 7~mm the degree of polarization rises along with 
the total flux density.    
An increase of both the degree of polarization and polarized flux density
as the VLBI core brightens implies that the flux increase is accompanied by
ordering of the magnetic field in the emission region.
The two opposing tendencies seen at 1~mm---lower fractional polarization
but higher polarized flux as the total flux becomes higher---indicate that the flux
increases as emission with weaker (but non-zero) polarization becomes prominent.

\section{Polarization Position Angle Behavior} 
\subsection{Comparison of Polarization Position Angle at Different Wavelengths}
\begin{figure}
\epsscale{1.0}
\plotone{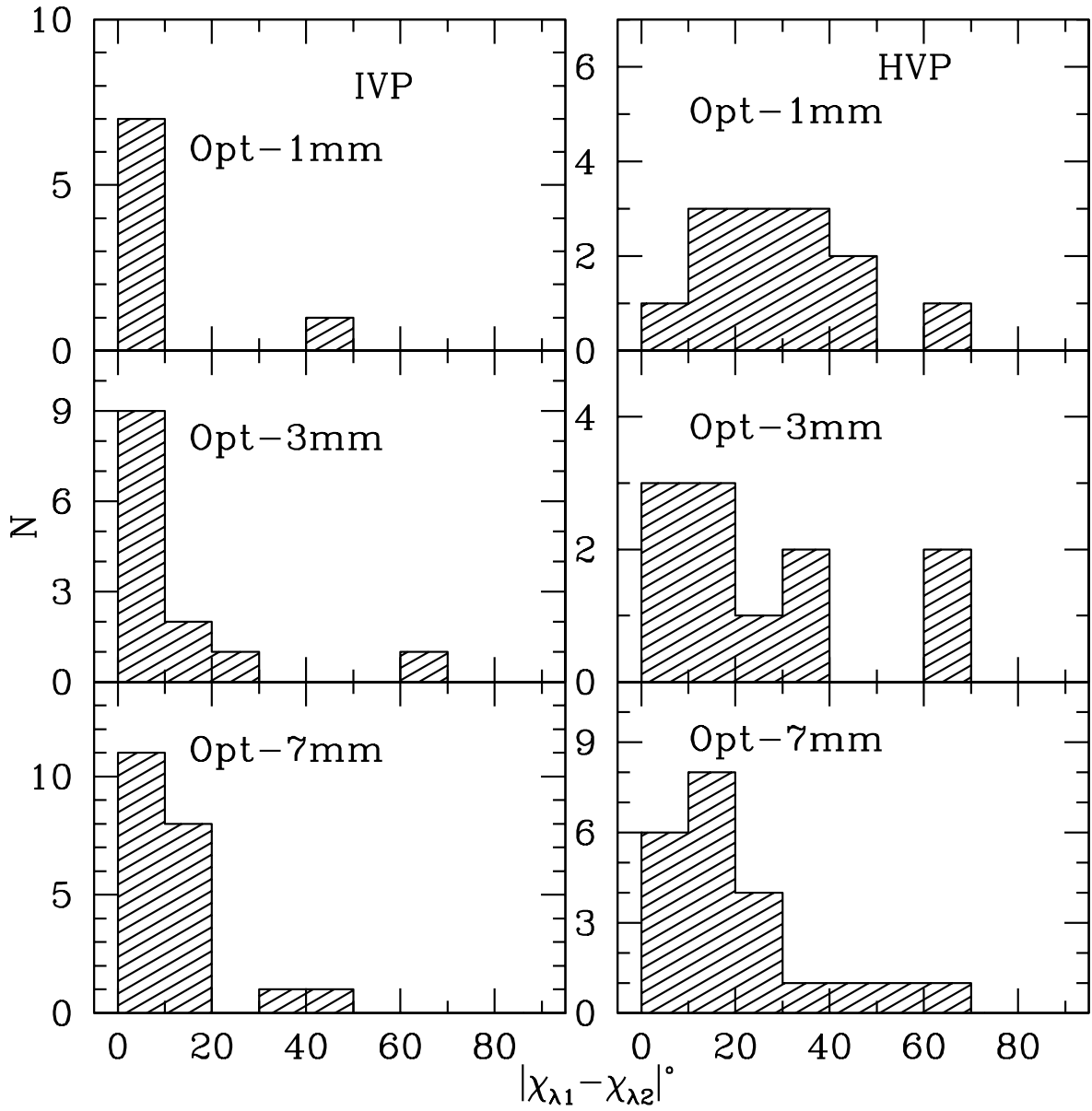}
\caption{Distributions of offsets between EVPAs
for simultaneous observations at different wavelengths
in the IVP and HVP sources.}  \label{hEVPA_EE1}
\end{figure}

For the IVP and HVP sources we have computed values of deviation between optical
and mm-wave EVPAs, $\Delta\chi_{\lambda_1\lambda_2}=|\chi_{\lambda_1}-\chi_{\lambda_2}|$, where $\lambda_1$ is the optical wavelength and 
$\lambda_2$ is 1~mm, 3~mm, or 7~mm, for measurements obtained within 2 
weeks of each other (recall that $\chi_{\rm 7mm}$ corresponds to the 
EVPA in the VLBI core). Figure \ref{hEVPA_EE1} presents
the distributions of $\Delta\chi_{\lambda_1\lambda_2}$.
  
Figure \ref{hEVPA_EE1} shows that the IVP sources possess excellent agreement
between polarization position angle at different wavelengths: 87\% of $\Delta\chi_{\rm o1}$ (8), 92\% of $\Delta\chi_{\rm o3}$ (12), and 90\% of $\Delta\chi_{\rm o7}$ (21) fall into the range 0$^\circ$--20$^\circ$
(the number in parentheses indicates the number of
observations). The distribution of $\Delta\chi_{\rm o7}$
for the IVP sources agrees very well with the distribution of
$|\chi_{\rm opt}-\chi_{\rm core,7mm}|$ obtained by \citet{GRSS06} for
11 BL~Lac objects and 3C~279. This suggests that in the majority of BL~Lac objects
(i) alignment between the optical polarization position
angle and EVPA in the VLBI core at 7~mm is a common feature, and 
(ii) good agreement of the EVPA at different wavelengths from optical to 
mm wavelengths is expected.
 
For the HVP sources, the result is
qualitatively different: 31\% of $\Delta\chi_{\rm o1}$ (13), 
55\% of $\Delta\chi_{\rm o3}$ (11), and 64\% of $\Delta\chi_{\rm o7}$ (22)
are located within 20$^\circ$. The best agreement is observed between optical 
EVPA and polarization position angle in the core at 7~mm, and between optical
and 3~mm EVPA, while the distribution of $\Delta\chi_{\rm o1}$ indicates
significantly larger misalignments between polarization position angles. 
    
\begin{figure}
\epsscale{1.0}
\plotone{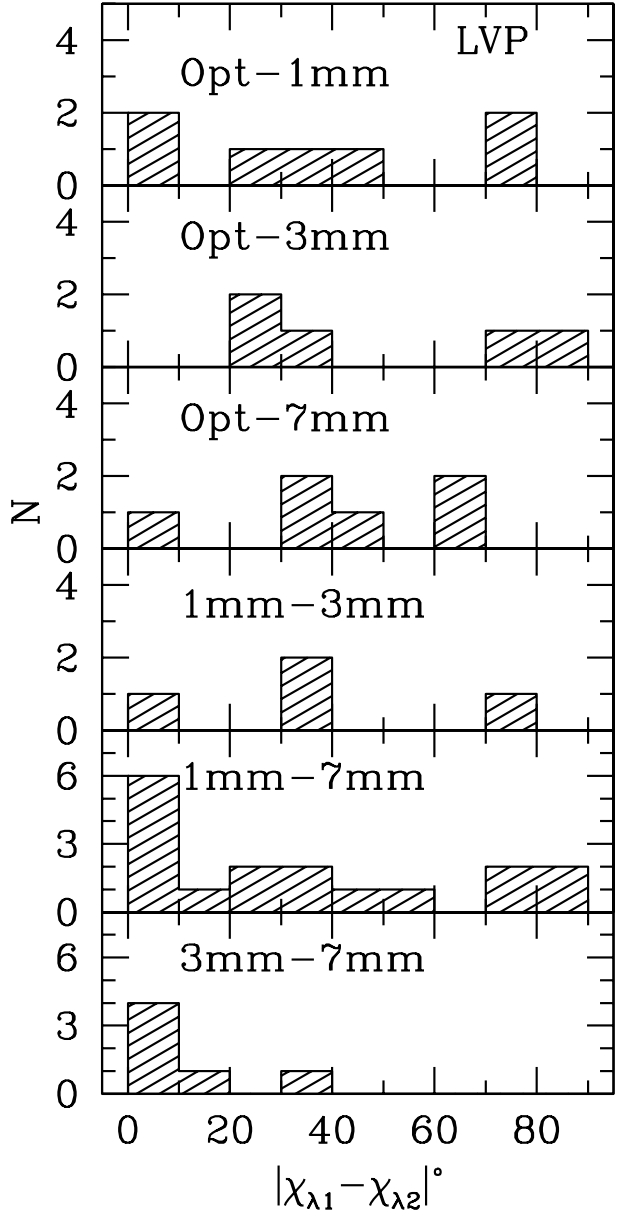}
\caption{Distributions of offsets between EVPAs
for simultaneous observations at different wavelengths
in the LVP. The polarization position angle at 7~mm corresponds
to the EVPA in the inner jet.}  \label{hEVPA_EE2}
\end{figure}

In the LVP sources the EVPA at 7~mm corresponds to $\chi_{\rm jet}$, 
defined by jet components within a few milliarcseconds of the core.
Since the EVPAs are not corrected for rotation measure, distributions   
of $\Delta\chi_{\lambda_1\lambda_2}$ are constructed for all possible pairs
of wavelengths (Fig. \ref{hEVPA_EE2}).
Figure \ref{hEVPA_EE2} shows that for the LVP sources the best agreement
in direction of polarization occurs between 1~mm and 7~mm
as well as between 3~mm and 7~mm. This result, plus a good correlation 
between the fractional polarization light curves at these wavelengths
found in \S 6.1, imply that in the LVP sources 
(i) the mm-wave core has weak polarization and (ii) high polarization at the short
mm wavelengths ($>$2-3\%) originates in a jet component within a few mas 
of the core. The findings require either (1) negligible 
Faraday depolarization at 7~mm and intrinsically low polarization in the core
at sub-mm wavelengths, or (2) a highly Faraday-thick screen near 
the core region, causing depolarization of the core at 1~mm. 
The latter is a more suitable explanation due to the high Faraday
rotation in the region nearest to the core, as discussed in \S 5.2. 
 
Comparison of EVPAs obtained 
at different wavelengths indicates that in the highly
optically polarized sources (the IVP and HVP blazars)
the optical polarization position angle tends to align with the EVPA
in the VLBI core, while the polarization
position angle at 1~mm in the HVP sources shows the weakest connection with
the optical polarization position angle.

\subsection{Direction of Polarization relative to the Jet Axis}
We have compared the polarization position angle, $\chi_\lambda$, 
for each polarization measurement with the position angle of the jet projected 
on the sky, $\Theta_{\rm jet}$ (Table \ref{TAB_VLBA}). $\Theta_{\rm jet}$
corresponds to the nearest VLBA epoch to a given polarization measurement.
Figure \ref{h_EVPA} gives the distributions of the offsets
between direction of polarization and jet axis, 
$|\chi_\lambda - \Theta_{\rm jet}|$,
for polarization measured at the different wavelengths, $\lambda$, and 
for the different groups of objects. In the case of the IVP and HVP
sources, $\chi_{\rm 7mm}$ corresponds to the EVPA in the core; 
in the LVP group $\chi_{\rm 7mm}$ is equal to $\chi_{\rm jet}$ 
as defined in \S 4. The EVPAs at 3~mm and 7~mm 
for the HVP and IVP sources are corrected for $RM$ and 
the distributions at these wavelengths should be similar to that
at 1~mm if the $RM$ variability is not significant. It is possible
though that the
difference in the number of observations at the mm wavelengths
can introduce some discrepancy between the distributions.

Figure \ref{h_EVPA} shows obvious alignment of the electric 
vector with the jet direction in the IVP sources: 75\% of $\chi_{\rm opt}$ (36),
59\% of $\chi_{\rm 1mm}$ (56), 76\% of $\chi_{\rm 3mm}$ (29), and
64\% of $\chi_{\rm 7mm}$ (102) lie within 20$^\circ$ of the inner jet
direction (the numbers in parentheses indicate the number of observations 
at each wavelength), although the widest range of offsets is observed
at 1~mm.   
In contrast, the HVP sources do not show any significant relation between
EVPA and jet direction, and the values of $\chi_\lambda$ are distributed 
almost uniformly from 0$^\circ$ to 90$^\circ$ with respect to the jet direction
at all wavelengths. An exception is the distribution at 1~mm, which contains 
two peaks at intervals 0$^\circ$--20$^\circ$ and 80$^\circ$--90$^\circ$, 
suggesting that there is a preference for the electric vector to lie along or
perpendicular to the jet. The difference between the distributions of the IVP and HVP
sources is similar to the difference between the distributions
of core electric vector offsets from the jet direction found by  
\citet{LH05} for BL~Lac objects and quasars, respectively, in the MOJAVE 
survey at 2~cm (Monitoring of Jets in AGN with VLBA Experiments). Although our dichotomy 
is not based simply on the optical 
classification of an object, we conclude that this difference persists from 
cm to optical wavelengths and implies differences in
the magnetic fields and/or processes responsible for the polarized emission 
in these objects.

In the LVP sources the distributions of $|\chi_\lambda - \Theta_{\rm jet}|$
at optical and 1~mm wavelengths do not support a connection of the EVPA with
the jet direction, although  the polarization at 1~mm seems to avoid being
parallel to the jet. The polarization position angle at 3~mm and in the inner jet
at 7~mm clearly reveals a preferential direction perpendicular to the jet:
89\% of $\chi_{\rm 3mm}$ (9) and 62\% of $\chi_{\rm 7mm}$ (47) values
lie within 70$^\circ$--90$^\circ$ of the jet axis. 
Although the distribution at 3~mm is dominated by the observations of 3C~273,
all three sources make a significant contribution in the pronounced
peak of the distribution at 7~mm.

The properties of polarization position angle with respect to the jet direction
are distinct for each group:
good alignment of the electric vector with the jet direction
at all wavelengths in the IVP sources, chaotic behavior of the electric
vector in the HVP blazars independent of frequency, and 
electric vector preferentially 
transverse to the jet direction in the LVP
objects at 3~mm and in the inner jet at 7~mm. 
It is noteworthy that the dichotomy at high frequencies is similar to that  
found for regions farther out in the jet at lower frequencies \citep{CAW93,LH05}.

\section{Interpretation}
\subsection{Magnetic Turbulence}
Our data are consistent with the assumption that 
we observe incoherent synchrotron radiation across the wavelength 
range from optical to mm wavelengths. A uniform magnetic field throughout
the emission region yields the maximum possible degree of polarization, 
$m_{\rm max}=(\alpha+1)/(\alpha+5/3)$ \citep[e.g., ][]{PAH70}, 
which is $\sim$70\% at mm wavelengths 
($\alpha\sim 0.5$, Table \ref{TAB_Spec}) and $\sim$78\% in the optical band
($\alpha\sim 1.4$; \citealt{IN88}). Such a high degree of polarization 
has never been observed in a blazar.  
The polarization can be significantly reduced if the emission region 
is turbulent so that it is composed of many cells, each containing a roughly
uniform magnetic field that
is randomly oriented relative to that in other cells \citep{BURN66, BURCH79}. 
For $N$ cells within a telescope beam, the fractional
polarization is given by the equation \citep{HM91}:
\begin{equation}
m\approx m_{\rm max}/\sqrt{N}\pm m_{\rm max}/\sqrt{2N}. \label{e3}
\end{equation}

The results obtained in \S\S 6-7 indicate that the polarization properties,
and therefore magnetic field structure, of our 
sources are comparable within a group but diverse across different groups in
our classification scheme.  
We have calculated for each
group the mean and standard deviation of the maximum degree of
polarization over sources measured at each wavelength, 
$\langle{m_\lambda^{\rm max}}\rangle\pm\sigma_\lambda^{\rm max}$, as well as 
similar values, $\langle{m_\lambda^{\rm min}}\rangle\pm\sigma_\lambda^{\rm min}$,
for the minimum degree of polarization. The values, given in Table 
\ref{TAB_POL}, show that both the maximum and minimum 
degree of polarization increase with frequency except for the LVP group, 
within which the optical polarization is similar to the polarization in the VLBI core.
At a given frequency, the fractional polarization is lowest for the
LVP objects and highest for the IVP sources, independent of whether the polarization
is near the minimum or maximum.
We use the means and standard deviations given in Table \ref{TAB_POL}  and
equation~(\ref{e3}) 
to determine whether the observed parameters can be derived within a cellular
model. Table \ref{TAB_POL} lists the number of cells,
$N^{\rm max}_{\rm cell}$ and $N^{\rm min}_{\rm cell}$, that result in the
observed degree of polarization under this model.

We can calculate the number of turbulent cells for any degree of polarization; however,
the same number should account for the standard deviation as well. We find that 
this requirement is not satisfied at high-polarization states, where the
standard deviation is significantly lower than the model predicts.
In low-polarization states,
the requirement holds at all wavelengths except 1~mm, where the
standard deviations for the LVP and HVP sources are slightly lower than
expected according to the derived number of cells.
This suggests that the high-polarization states involve an additional 
process that provides some order to
the magnetic field, while the low-polarization states can be explained within the
cellular model. 

Three interesting consequences follow from analysis of Table \ref{TAB_POL}.
First, the observed difference in degree of polarization
among the groups of objects suggests 
that the emission regions of the LVP and HVP sources
contain more turbulent cells than is the case for IVP objects -- either the
cells are finer in the LVP and HVP sources, or the emission region is smaller
in the IVP blazars. 
Second, the fact that sources within a group have similar fractional polarization
at different wavelengths implies that the emitting regions partially overlap.
Third, the fact that fractional polarization decreases with wavelength
suggests that the emitting region is larger at longer wavelengths.

\begin{figure}
\epsscale{1.0}
\plotone{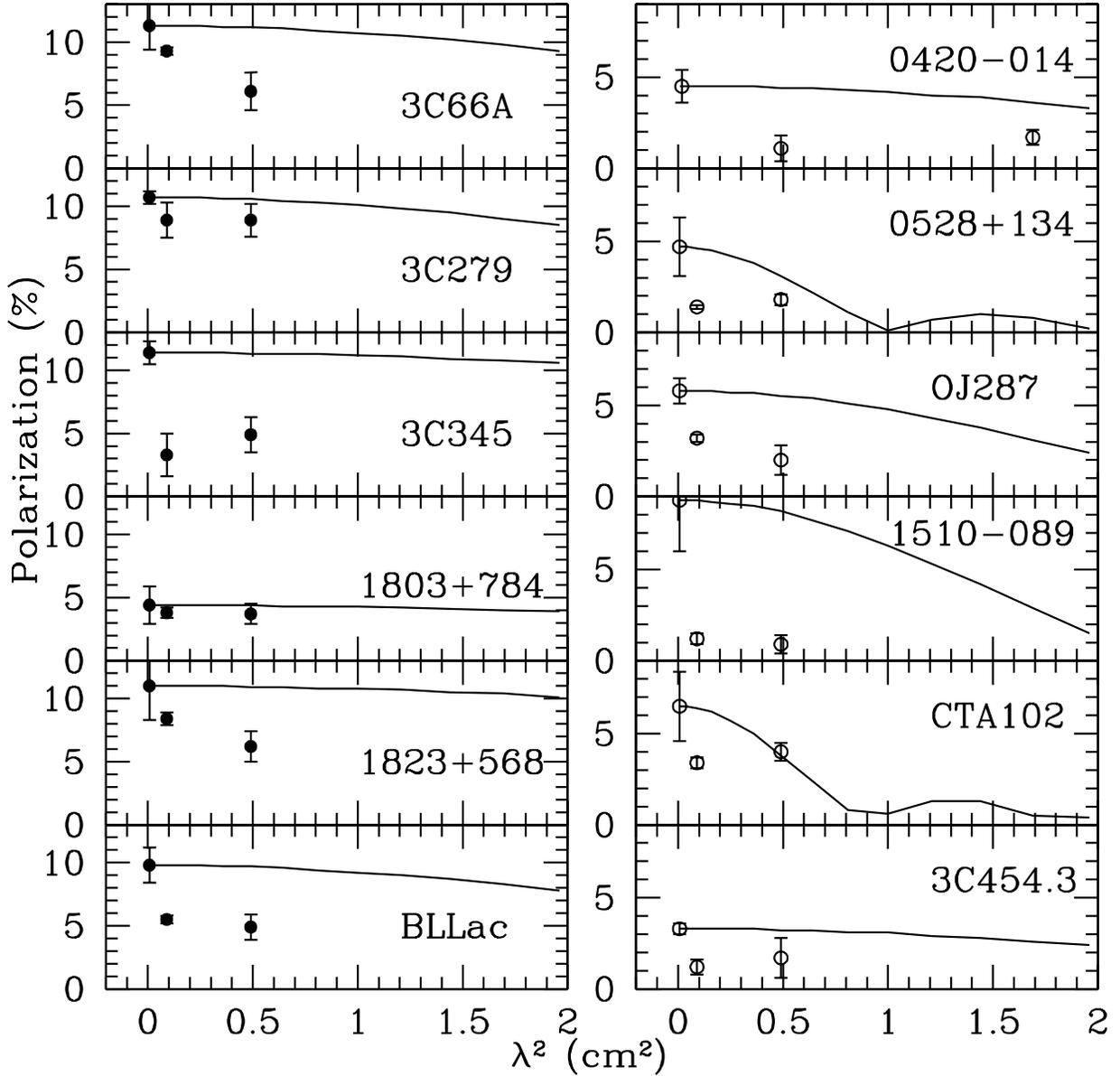}
\caption{Dependence of degree of polarization on wavelength in the IVP
({\rm left panel}) and  HVP ({\rm right panel}) sources.
Solid lines represent the 
expected Faraday depolarization according to eq.~\ref{e4} and
$RM$ derived in \S 5.1. In general, the fit is poor, so that
in most of our sources Faraday depolarization is unable to explain
the frequency dependence of the polarization.} \label{RM_depol}

\clearpage
\end{figure}

Faraday rotation by a foreground screen can produce 
depolarization across a beam that also results in frequency 
dependence of the polarization \citep{BURN66}. However, Faraday 
depolarization can not account
for the frequency dependence among the groups
at optical and 1~mm wavelengths since the effect is negligible
at such short wavelengths. We assume that turbulence and difference
in volume of emission is
responsible for the frequency dependence at optical and 1~mm
wavelengths but the decrease of polarization at longer mm-waves 
could be caused by Faraday depolarization. To calculate the 
frequency dependence of the polarization we apply a model proposed 
by \citet{BURN66}, with the simplifications suggested by \citet{ZT04}:
\begin{equation}
m(\%) = m_{\rm RM=0}|\rm{sinc}(RM\lambda^2)|,\label{e4}
\end{equation}
where $RM$ values are rotation measures derived in \S 5.1.
We compare the model with the polarization 
at different wavelengths using the simultaneous data employed for $RM$ estimates.
In this case $m_{\rm RM=0}$ corresponds to $m_{\rm 1mm}$  
and is different for each source. 
Figure \ref{RM_depol} shows the frequency dependence of the polarization
according to the observations and eq.~(\ref{e4}).
Figure \ref{RM_depol} demonstrates that although for some sources
(3C~279 and 1803+784) the dependence of the polarization
on frequency can be caused by Faraday depolarization, for the majority
of the objects the fractional polarization decreases with wavelength
much faster than eq.~(\ref{e4}) predicts. It appears that depolarization from 
a gradient across the beam in a Faraday screen alone is unable to explain  
the observed frequency dependence of the polarization. A difference in the size
of the emitting regions and/or in turbulence scale of the magnetic field
is needed to account for the dependence. 

\subsection{Shock Model vs. the Observed Polarization}
A commonly occurring structure that leads to ordering of magnetic fields is a shock,
which compresses the component of the magnetic field that lies
parallel to the shock front. The significant dichotomy in degree of polarization
at high-polarization states among the groups might then be explained by 
differences in shock strengths. \citet{LS00} proposed a similar explanation
to account for differences in polarization properties of low and high
optically polarized radio-loud quasars (LPRQs and HPQs, respectively),  
suggesting that LPRQs have weaker shocks than do HPQs. 
We use the \citet{HM91} approach to estimate the shock strength for the sources
in our sample: 
\begin{equation}
m\approx \frac{\alpha+1}{\alpha+5/3}\times\frac{(1-\eta^{-2})\sin^2\Theta'}{2-(1-\eta^{-2})\sin^2\Theta'}, \label{e5}
\end{equation}
where $\eta=n_{\rm shocked}/n_{\rm unshocked}$ and $n$ is the number density of the plasma; 
$\Theta'$ is the viewing angle of the jet corrected for relativistic aberration: 
$\Theta'=\tan^{-1}[\sin\Theta_\circ/(\Gamma(\cos\Theta_\circ-\sqrt{1-\Gamma^{-2}}))]$,
where $\Gamma$ is the bulk Lorentz factor and $\Theta_\circ$ is the viewing angle of the
jet in the observer's frame. The use of the viewing angle with respect to the jet axis 
in the formula applies to a
transverse plane-wave shock, and the formula is valid for a shock propagating through
a plasma with a completely chaotic magnetic field. The maximum possible polarization should
be seen when the shock is viewed along the plane of compression
in the frame of the shock (i.e., $\Theta' \approx 90^\circ$). 
Columns 3, 5, 7, and 9 of Table \ref{TAB_Shock} list the values of the estimated shock
strength $\eta$ based on the maximum observed degree of polarization (columns 2, 4, 6,
and 8) at each wavelength. The last column of the table
contains the viewing angle $\Theta'$ in the shock frame,  as derived from the values
of $\langle\Theta_\circ\rangle$ and $\langle\Gamma\rangle$ obtained for each object in J05. 
Because we do not have
multi-color data in the optical region, the computation
of $\eta$ adopts $\alpha_{\rm opt/1mm}$ for the value of $\alpha_{\rm opt}$, while
at 1~mm, 3~mm, and 7~mm we use the three-point spectral index $\alpha_{\rm mm}$ derived
from all three mm wavelengths (see Table \ref{TAB_Spec}).

The shock strengths listed in Table \ref{TAB_Shock} are similar among the
sources and rather low, $\eta\sim 1.1-1.2$, and hence cannot be the primary reason
behind the diversity in degree of polarization. Instead, we can explain this by
differences in viewing angle relative to the plane of compression, although
a very high level of polarization ($m\gtrsim$30\%) does
require much stronger compression, $\eta\sim 2-3$. Note that for 3C~66A 
J05 find a significant range of apparent speeds
of superluminal components, from 1.5$c$ to 27$c$. We use the Lorentz factor and 
viewing angle that corresponds to the average of $\Gamma$ and $\Theta_\circ$ 
derived for the fast components only ($\Gamma$=27.8 and 
$\Theta_\circ$=3.3$^\circ$),
because $\langle\Gamma\rangle$ and $\langle\Theta_\circ\rangle$ averaged
over all components (J05) fails to produce
polarization as large as that observed for any given value of $\eta$. 
The latter is the case for 1823+568 as well. However, for 1823+568 
moving components within 1~mas of the core were not observed (J05)
and the jet parameters are obtained
using kinematics of two components at $\sim$2~mas from the core.
This implies that $\Gamma$ and $\Theta_\circ$ in the VLBI core of 1823+568 
might be different from those derived in J05. 

\begin{figure}
\epsscale{1.0}
\plotone{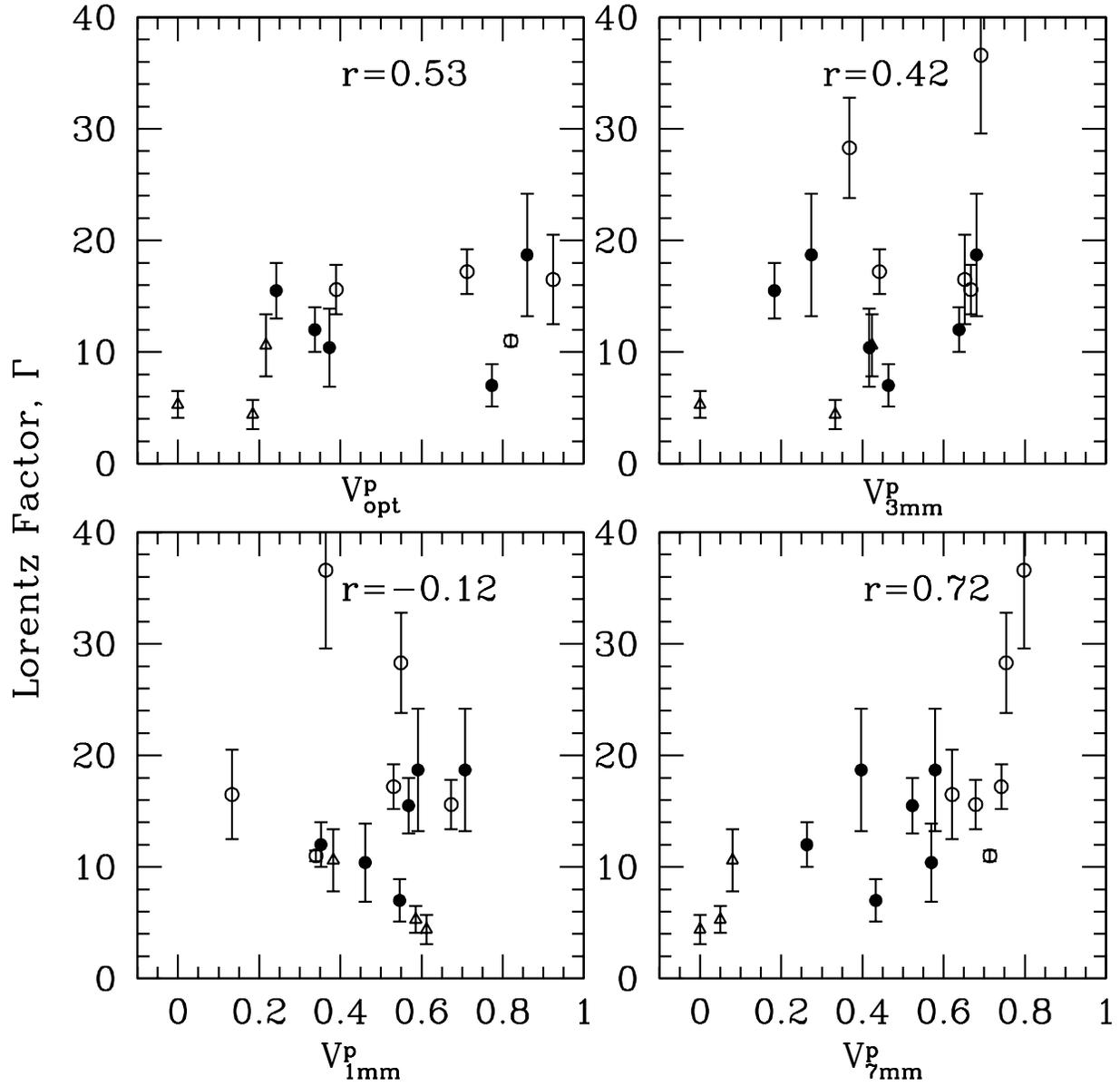}
\caption{Lorentz factor vs. polarization variability index at different 
wavelengths. Symbols have the following meaning: triangles - LVP sources, filled circles - IVP
sources, open circles - HVP sources.} \label{G_Vp}
\end{figure}

Figure \ref{G_Vp} shows the dependence
between bulk Lorentz factor and polarization variability index.
There is a correlation between $\Gamma$ and $V^{\rm p}_{\rm opt}$ 
($r$=0.53, $\epsilon=0.1$) and
between $\Gamma$ and $V^{\rm p}_{\rm 7mm}$ ($r$=0.72, $\epsilon=0.05$).
Although $V^{\rm p}_{\rm 3mm}$ correlates with $\Gamma$ ($r$=0.42), 
this correlation is significant only at a level $\epsilon >$0.15.
In the plane-wave shock model, the correlation is expected if the minimum degree 
of polarization corresponds to unshocked plasma and the maximum to faster, shocked
plasma. In the shock-in-jet model, polarized regions at different wavelengths are
partially co-spatial and have essentially the same bulk Lorentz factor. Radiative
energy losses of the relativistic electrons cause the shorter wavelength emission
to occupy a smaller fraction of the full shocked region seen at longer wavelengths
\citep[][]{MG85,MAR96}. Since the shorter wavelength emission is assumed to originate from fewer
turbulent cells, the degree of polarization and its variability should increase with 
decreasing wavelength according to equation (\ref{e3}).
 
Contrary to what we find at the other wavelengths, the polarization variability 
index at 1~mm does not correlate
with the Lorentz factor ($r$=$-$0.12; see Fig. \ref{G_Vp}). 
Furthermore, according to Table \ref{TAB_POL}, the number of cells needed in the 
high-polarization state at 1~mm is only slightly larger than, or comparable to, that 
in the optical region. This conflicts with the expectations of the shock model, although
$N^{\rm max}_{\rm cell}$ at 1~mm is much smaller than  at 3~mm and 7~mm, in keeping
with the model. The partial discrepancy suggests that another emission component in addition 
to transverse shocks is prominent at 1~mm. A primary candidate is the
``true'' core, i.e., the bright, narrow end of the jet when observed at a wavelength
where the emission is completely optically thin (Fig. \ref{Sketch}). At wavelengths of 3 mm and longer,
what appears to be the core is most likely a location at or outside the point
where the optical depth
$\tau(\nu)\sim 1$. The fact that the optical polarization 
does show the correlation, as well as the best alignment of 
position angle with the electric vector in the core at 7~mm,
implies that most of the nonthermal optical emission arises in shocks close
to the 7~mm core rather than in the``true'' core. This suggests that the
``true'' core does not possess relativistic electrons energetic enough 
to produce the optical synchrotron emission.

\subsection{Magnetic Field Structure in the Jets} 
In the shock-in-jet model \citep[e.g.,][]{MG85,HAA85}, we expect
a correlation between total flux and polarization owing to ordering of 
the magnetic field in the shocked region if the quiescent jet has a completely chaotic 
magnetic field. However, if the magnetic field in the unshocked 
region possesses a component parallel to the jet axis, a transverse shock will enhance
the total flux but partially cancel the polarization as it compresses the turbulent
component. In this case the polarized flux will at first
decrease as the total flux rises. If the shocked emission grows strong enough that its
polarized flux becomes greater than the initial polarized flux of the ambient jet,
the EVPA will flip by $90^\circ$ after a minimum in polarized flux, and the polarized flux 
will subsequently grow with the total
flux until the shocked emission weakens. 
The initial phase of decline in polarized flux may easily 
be missed in observations spaced such as ours.  
If an oblique shock is responsible for an outburst, its effect on the polarization
will depend on the angle of the shock front relative to the ordered component
of the ambient magnetic field, the viewing angle, and the bulk Lorentz factor of
the jet. It is possible, in some cases, for the net
polarization to increase, with the EVPA becoming closer to the position angle of the
jet axis on the sky \citep{PH05}.

\begin{figure}
\epsscale{1.0}
\plotone{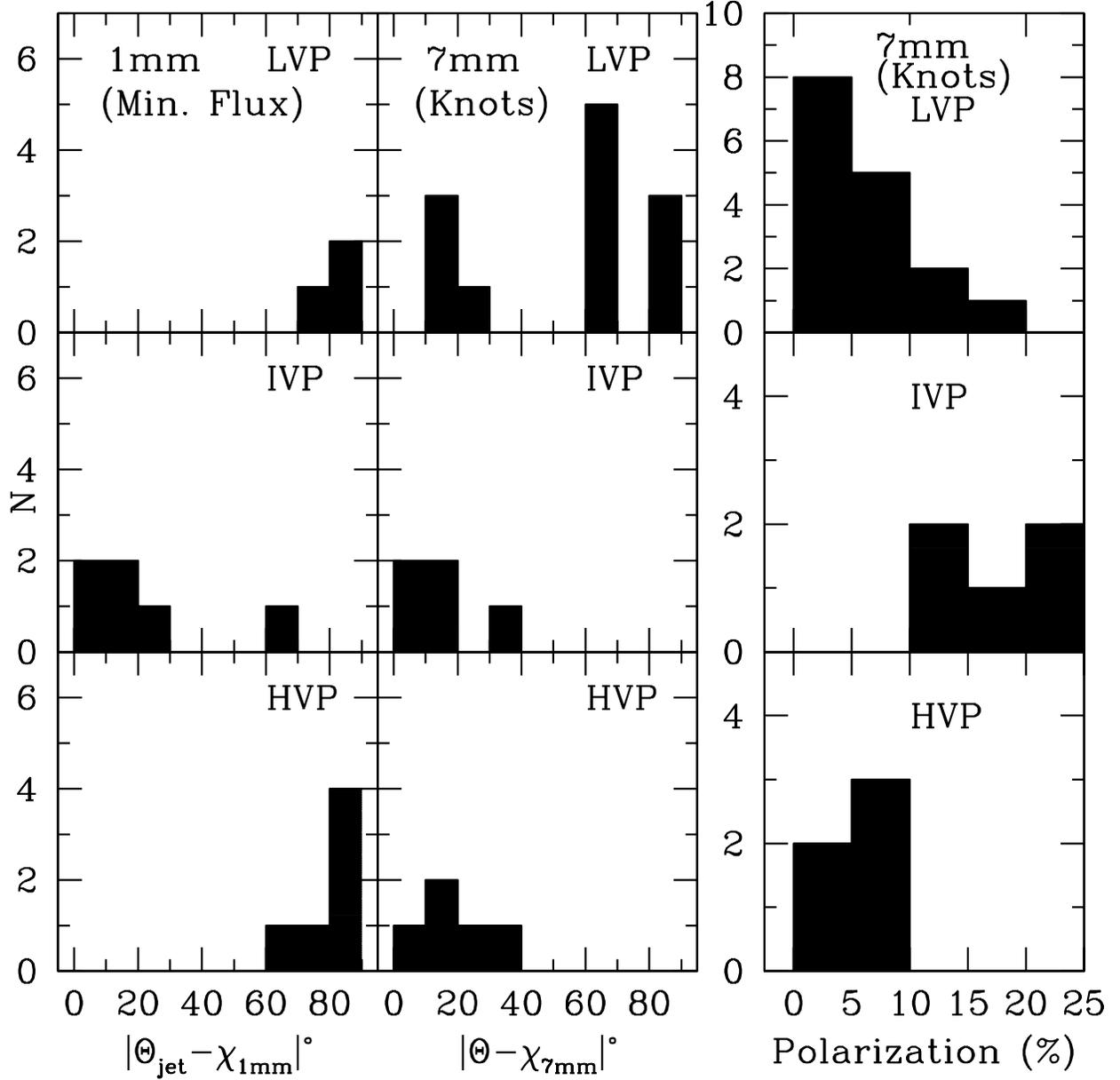}
\caption{{\it Left panel:} Distributions of offsets between polarization position
angle at 1~mm and inner jet at minimum flux observed 
at 1~mm in each source. {\it Middle panel:} Distributions of offsets between
polarization position angle at 7~mm and local jet direction for 
superluminal knots listed in Table \ref{TAB_Pcomp}.
{\it Right panel:} Distributions of the degree of polarization in superluminal
knots (one entry for each knot; see text).} 
\label{h_shock}
\end{figure}

Figure \ref{FPOL} shows that the maximum of degree of polarization in 
the core at 7~mm tends to occur when the total flux rises, while
at 1~mm the opposite relation is more probable.
An increase of the total flux in the VLBI core is usually connected
with the emergence of a new superluminal component from the optically thick part of the core (see J05).  
If these components represent transverse shock formation in the jet, 
they should display electric vectors aligned
with the jet axis. On the other hand, $\chi_{\rm 1mm}$ at minimum 
flux might reflect the magnetic field direction in the unshocked region. 
Figure \ref{h_shock} ({\it left panel}) displays
the distributions of misalignment angles between the polarization  
and inner jet directions at times of minimum observed flux at 1~mm.
Figure \ref{h_shock} ({\it middle panel}) shows the distributions of 
offsets between $\chi_{\rm 7mm}$ and the local jet direction 
for each superluminal component
listed in Table \ref{TAB_Pcomp} at the point when we can first separate the knot
unambiguously from the core on the image (i.e., farther downstream than 0.1~mas).  
Figure \ref{h_shock} ({\it right panel}) presents the distribution of degree 
of polarization in the superluminal knots at their maximum total flux when 
the separation from the core $>$0.1~mas. Figure \ref{h_shock} indicates
that (1) in the ``quiescent jet'' (prominent at minimum 1~mm flux), the magnetic 
field aligns with the jet for the LVP and HVP sources but is transverse to the jet 
in the IVP blazars; (2) in the shocked regions 
(superluminal knots) of the LVP objects, the magnetic field is longitudinal 
or oblique to the jet, while for the IVP and HVP sources, its apparent direction is
perpendicular to the jet. 

\begin{figure}
\epsscale{1.0}
\plotone{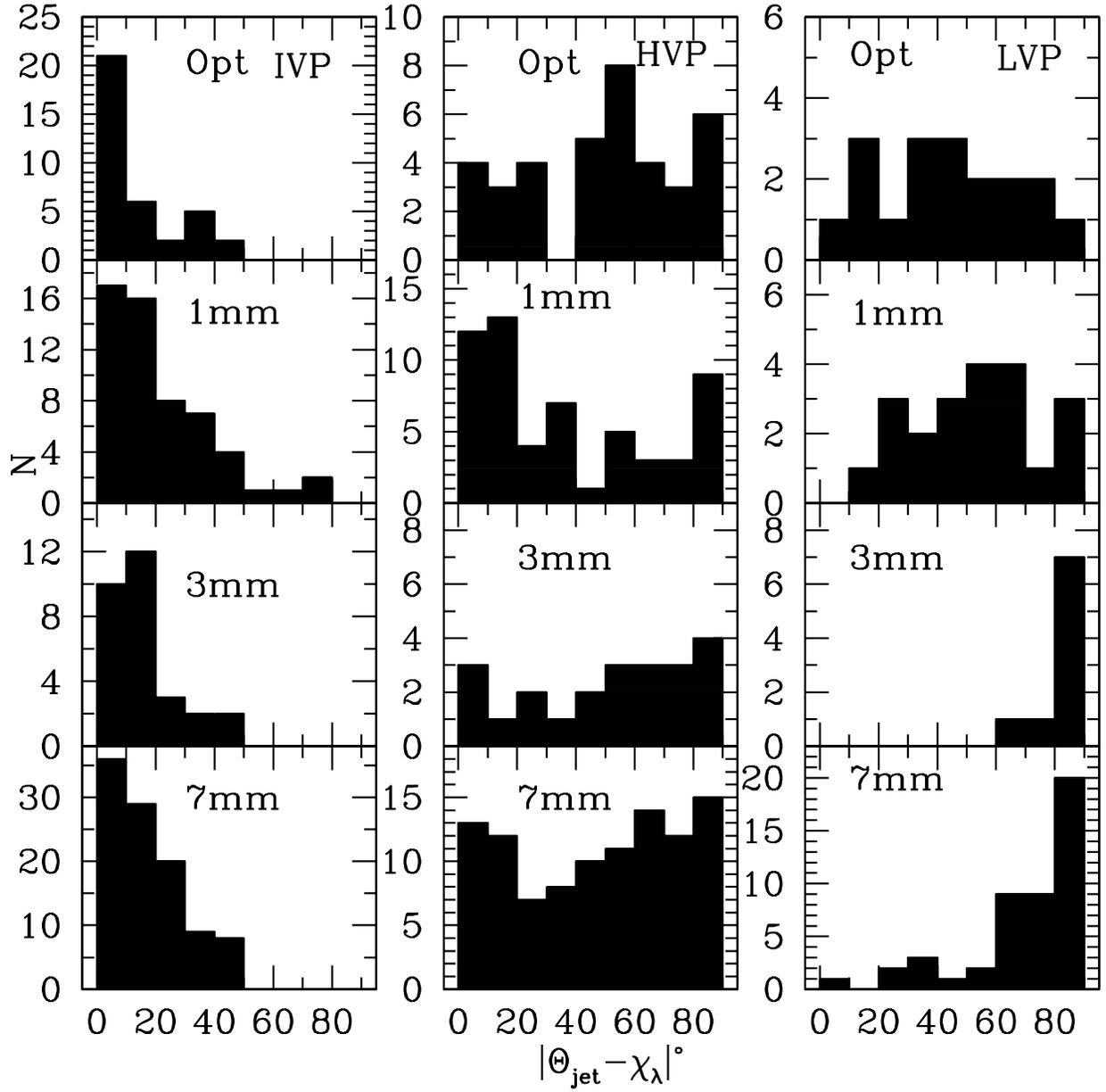}
\caption{Distributions of offsets between polarization position angle 
and jet direction.}  \label{h_EVPA}
\clearpage
\end{figure}

The distributions of offsets between the electric vector and jet 
direction obtained for the IVP sources at different wavelengths (Fig. \ref{h_EVPA}),
the distribution of EVPAs in the superluminal knots,
and the high polarization measured in the knots agree with the case
of magnetic fields lying at oblique angles to the jet axis 
\citep{LMG98}. (Relativistic aberration tends to align 
the polarization vectors closer to the jet axis, creating the illusion that 
the magnetic fields are closer to transverse
than is actually the case in the flow frame of the emitting plasma.)
This result matches the finding for BL~Lac
objects reported by \citet{ALL03b} and \citet{ALL05} that both the linear polarization 
structure in VLBA images and the
multifrequency polarized flux light curves obtained with 
single-dish observations are consistent with the oblique shock hypothesis.
Note that the IVP sources include most of the BL~Lac objects plus those quasars with 
similar VLBI polarization structure.

In the HVP sources the distribution of offsets between the electric
vector and jet direction at 1~mm is double-peaked 
(Fig. \ref{h_EVPA}) -- nearly parallel or perpendicular to the jet. 
According to Figure \ref{h_shock}, at minimum flux at 1~mm the magnetic field has 
a parallel orientation, which could be associated with the unshocked plasma.
A longitudinal component of the magnetic field in a quiescent jet can arise 
from a velocity shear across the jet flow \citep{Laing80}. 
Subtraction of the distribution shown in Figure \ref{h_shock} 
from that plotted in Figure \ref{h_EVPA} 
produces a distribution for the HVP sources at 1~mm that is quite similar to the distribution of 
the IVP sources. Therefore, the magnetic field in the regions responsible for 
enhanced emission, including superluminal knots, lies at a transverse or 
oblique angle to the jet axis, in agreement with the shock model.
The superluminal components in the HVP sources (Fig. \ref{h_shock})
have only a moderate degree of polarization ($<$10\%), indicating that
even in the shocked region, the magnetic field is fairly turbulent. This is 
consistent with the high amplitude and short timescale of the polarization 
variability seen at all wavelengths. The existence of a longitudinal component 
of the magnetic field in quiescent 
states suggests that either an increase or a 
decrease in the degree of polarization is possible when a source brightens, 
depending on the shock strength responsible for the flux enhancement. 
This diversity of behavior reduces the possibility of observing a correlation
between total flux and fractional polarization. 

The general misalignment between the inner jet
and the polarization vector in the core at 7~mm and at optical and 3~mm 
wavelengths for the HVP blazars (Fig. \ref{h_EVPA}) fails to reveal
any connection between the magnetic field and jet structure.
In the shock model the distribution of electric vector offsets
is expected to show either a good alignment with the jet direction
(transverse or oblique shock, \citealt{HAA85,LMG98}) or a bimodal
shape (conical shock, \citealt{CC90,CAW06}). In the conical shock wave
model, the on-axis polarization in the shocked region
corresponds to the upstream magnetic field structure, while the off-axis 
electric field vectors are oblique to the axis and cover a wide range
of angles that depend on the shock cone opening angle and the viewing angle.
A slight asymmetry of the jet cross-sectional structure 
with respect to the axis should lead to 
relaxation of the dependence of the magnetic field on the jet direction.
Another factor that can contribute to the spread of $\chi$ with respect to the
jet axis is that, despite the high resolution of the 7~mm images, the jet
direction at the position where the emission occurs might be different from 
the direction inferred from the images. The latter implies that
HVP sources possess higher bending of the jets on milliarcsecond 
scales than do IVP sources. Conical shocks might explain the appearance 
of stationary features in the jet \citep{GOM97, DM88}, which are common 
in the HVP sources (J05). This model suggests that the core at 7~mm does not always
represent the emission from the surface where the optical depth $\tau$
is of order unity. Rather, at any given frequency below the turnover in the
spectrum, it would be either the $\tau \sim 1$ surface or
the first standing conical shock system in the $\tau < 1$ section of the
jet, whichever is stronger. If so, the model would explain
why the separation between the core and closest superluminal knot
has a frequency dependence in some sources and not in others
\citep[see][ and references therein]{MAR06}. 

The distributions of position angle offsets between electric vectors and jet
direction in the LVP sources (Figs. \ref{h_EVPA}, \ref{h_shock}) indicate 
the existence of a persistent longitudinal component of the magnetic field in 
the emission regions at 3~mm, 7~mm, in the quiescent jet, and in 
shocked regions.
The modest degree of polarization observed in superluminal components of 
the LVP sources requires a very turbulent magnetic field  
even in the shocked region. J05 found that LVP
sources have the widest jet opening angle
in our sample, and the quasar 3C~273 shows significant variation in the
apparent speed of superluminal components across the jet. 
The southern jet components are faster than the northern 
ones by a factor of 3. This suggests that a longitudinal component
of the magnetic field might arise from velocity shear in the ambient jet, or from
complex evolution of the shock emission owing to
the formation of oblique shocks combined with the effects of radiative transfer. 
The latter is seen in the numerical simulations of \citet{PH05}.

Using the parameters of jet components $B1$ and $B2$
in the quasar 3C~273 derived in J05 and the polarization information given
in Table \ref{TAB_Pcomp}, we estimate the distance from the core, $\Delta z$,
where a degree of polarization $p$ will be obtained as the result of
velocity shear in the jet flow. We assume that the magnetic field starts out in a 
turbulent state with cell size $\Delta x$:
 $\Delta z/\Delta R\approx\Gamma_{\rm N}\Delta R~p/(\Delta x\beta_{\rm rel})$,
where $\Gamma_{\rm N}$ is the bulk Lorentz factor at the northern side of the jet, 
$\Delta R$ is the half-width of the jet, and $\beta_{\rm rel}$ is the
relative speed between the northern and southern sides. In the plasma frame on the 
northern side (subscript ``$\rm{N}$''),
$\beta_{\rm rel}=(\beta_{\rm S}-\beta_{\rm N})/(1-\beta_{\rm N}\beta_{\rm S})$,
where $\beta=\sqrt{1-\Gamma^{-2}}$.
We assume that the Lorentz factor of knot $B1$ corresponds to the speed at
the northern side ($\Gamma_{\rm N}=8.3$) and the Lorentz factor of knot $B2$ gives
the speed on the southern side ($\Gamma_{\rm S}=13.8$). According to
Table \ref{TAB_Pcomp}, both components possess a similar maximum 
polarization ($p_{\rm B1}$=4.0\%,
$p_{\rm B2}$=6.2\%) when the direction of the electric vector
is transverse to the jet axis, so we approximate $p\sim$5\%.
Note that this polarization is achieved in both knots at the same 
distance from the core, $\sim$0.8~mas. We use the half-opening and 
viewing angle of the jet derived in J05 
($\theta=1.4^\circ$ and $\Theta_\circ=6.1^\circ$,
respectively). This exercise gives a ratio $\Delta R/\Delta x\sim$40 and 
size of a turbulent cell $\Delta x \sim 0.001$~pc.
For the quasar 3C~273 we have obtained several groups of closely spaced observations
at 3~mm near some VLBA epochs (see Table \ref{TAB_BIMA}). 
The data reveal a very short timescale 
\citep[computed according to the definition of][]{BJO74} 
of the fractional polarization variability at 3~mm, 
$\Delta t_{\rm var}\sim 1.8$~day. 
This yields the size of the region responsible for the variability, 
$\Delta x_{\rm var}\sim 0.001$~pc, which is consistent with the scale 
of turbulence in the magnetic field. 

In summary, the sources in our sample show significant variety in the distributions 
of offsets between polarization position angle and jet direction. 
The observed behavior can be explained within the context of shock waves 
compressing highly turbulent 
magnetic fields and velocity shear stretching field lines along a
direction parallel to the axis if the jet structure and kinematics on milliarcsecond 
scales are taken into account.
 
\subsection{Possibility of an Electromagnetically Dominated Core at 1~mm}
The polarized emission at 1~mm possesses properties distinguished
from the other wavelengths: 
1) there is no correlation between fractional polarization variability
index and Lorentz factor of the jet;
2) the maximum degree of polarization
tends to occur when the total flux is lower than average;
3) the standard deviation 
of the mean of the minimum degrees of polarization over sources in a 
variability group is lower than expected in the cellular model of magnetic 
field structure.
An inverse relationship between the degree of polarization and total flux is 
expected in electromagnetically dominated (ED) jets \citep{LR03} 
if the plasma responsible for a rise in the total flux has a 
more chaotic magnetic field than that of the ED section. 
This can be the case even if a shock orders an underlying turbulent field
in the same sense as does the mostly toroidal field in the ED portion of the jet,
since the polarization of the ED region should be quite high. 

The upstream ED model of the jet is a promising scenario
since it explains the high Lorentz factors of the flow needed 
to produce the high apparent speeds seen in some blazars. 
Acceleration of the jet to such high values of $\Gamma$ is accomplished
by conversion of Poynting flux to flow
energy through the Lorentz force that occurs on
scales that are much larger than the gravitational radius of the
central black hole \citep[e.g.,][]{VLA06}. 
The tentative anticorrelation 
between total flux and degree of polarization at 1~mm and rather
positive correlation between the values 
at 3~mm and in the core at 7~mm suggest that the acceleration might end between
the VLBI cores at 1~mm and 3~mm. In this case the 1~mm emission region
should possess a large scale, mostly toroidal, magnetic field component, which in the plane 
of the sky should lie either transverse or longitudinal
to the jet direction \citep{LPG05}. This is consistent with the distributions
of EVPAs at 1~mm with respect to the jet direction in the IVP and HVP sources,
in which emission from the VLBI core at 1~mm seems to dominate the overall 
source emission at 1~mm (Fig. \ref{h_EVPA}). If
the section of the jet seen at 3~mm and 7~mm is beyond the
transition point where the jet becomes mainly hydrodynamical and turbulent, then
the polarization behavior at these wavelengths will be similar to that discussed 
in \S\S 8.1-8.3, with
shocks partially ordering the randomly-oriented component of the magnetic field. 
The behavior of the
optical polarization is more similar to that at 3~mm and 7~mm, and the 
best agreement between polarization position angles for simultaneous
observations is observed between the optical EVPA and polarization
position angle in the core at 7~mm. This suggests that
the optical emission originates in shocks at or downstream of the 
transition point, e.g., at or between the VLBI cores at 3~mm and 7~mm, and 
implies that high-energy
particle acceleration is more efficient in shocks than in the ED region.

\section{Summary}
We have performed multi-frequency, multi-epoch, quasi-simultaneous
linear polarization observations of a sample of AGN that possess highly
relativistic jets and high levels of polarization at 1~mm. We separate
the sample into three groups according to the pattern of
variability of the polarization in the VLBI core at 7~mm. This classification
appears to be connected with physical differences between the objects.
However, the sample is small and future observations of a larger
sample are needed to confirm the findings. The properties of the groups 
are as follows:

1) The objects with low variability of polarization (LVP) in the VLBI core 
unite two radio galaxies with superluminal speeds on parsec scales 
and a quasar with low observed optical polarization. The LVP sources have
measured optical polarization $<$2\% and radio core polarization $<$1\%, and 
produce strongly polarized radio jet components that are 
responsible for the high polarization observed at 1~mm. 
The LVP inner jets are most likely subjected to very high rotation measures,
$RM > 5\times 10^5$~rad~m$^{-2}$, that cause depolarization of the core region 
at mm wavelengths. The optical nonthermal emission is greatly diluted by 
non-synchrotron components, probably associated with the Big Blue Bump.
The parsec-scale jets are characterized by moderate Lorentz factors, near
the low end of the blazar range, and by viewing and opening angles that are
larger than those of a typical blazar. The distributions of EVPAs at 3~mm
and in the inner jet at 7~mm with respect to the jet axis reveal a preferred
longitudinal component of the magnetic field that can be explained by
velocity shear. 

2) The sources with intermediate variability of polarization (IVP) in the 
VLBI core include BL~Lac objects and quasars with preferred direction of 
the electric vector at all wavelengths close to the direction of the inner jet.
They possess very high optical polarization ($\gtrsim$30\%) and relatively 
highly polarized ($>$2\%) VLBI cores at 7~mm. The polarized and total flux density
spectra from optical wavelengths to 1~mm suggest that
synchrotron components with fairly flat spectra, $\alpha\sim$0.5,
dominate the total emission at these wavelengths.
The IVP sources possess a moderate rotation measure, $RM\lesssim 5\times 10^3$~rad~m$^{-2}$,
in the core that is consistent with the lack of emission lines in
the spectra of BL~Lac objects owing to a relatively low column density of gas. 
The radio jets of the IVP sources are highly relativistic, with $\Gamma\ge$10.
The alignment of the polarization with the jet at all wavelengths 
indicates the dominance of a transverse or (in the plasma frame) oblique
component of the magnetic field.

3) The category of objects with highly variable polarization (HVP) in the VLBI core
consists of the OVV quasars and the BL~Lac object OJ~287.
These display strong variations in both degree and polarization position angle.
The very high optical polarization seen in the IVP sources occurs
in the HVP sources, but, in contrast with the IVP blazars, low polarization 
($<$1\%) at mm wavelengths and in the VLBI core is common.
Only a single synchrotron component with a steep spectral index, $\alpha\sim$1.0,
is required to produce the polarized emission from optical to 1~mm wavelengths,
although  a non-synchrotron component could contribute a small fraction of the total
optical emission. The HVP sources possess high rotation measures in the VLBI
core at 7~mm, $RM\sim 3\times 10^4$~rad~m$^{-2}$, implying that the core 
is surrounded by dense thermal plasma. The radio jets of the HVP sources
have Lorentz factors at the high-end tail of the blazar distribution. 
The general misalignment between the polarization position angle and jet direction
indicates that there is little, if any, connection between the two. 

The diversity of polarization properties of the sources in the sample
allows us to establish connections between parameters at different 
wavelengths.
(i) There is good agreement between the optical EVPA and that in the VLBI core 
at 7~mm for the IVP and HVP blazars.
(ii) The overall polarization of a source at 1~mm and 3~mm strongly correlates
with the degree of polarization in the inner jet of the LVP sources. 
(iii) The EVPA at 1~mm and 3~mm corresponds to that in the inner jet for the LVP sources.
(iv) There is a strong correlation between the level of polarization
variability in the optical region and in the VLBI core at 7~mm.
(v) The optical and 7~mm polarization variability indices 
correlate with the bulk Lorentz factor of the radio jets.
(vi) The degree of polarization decreases with wavelength.
(vii) The distributions of electric vector offsets with respect to the  
inner jet direction at optical and mm wavelengths are similar within 
each variability group, with distinct differences among the three groups.

In general, the time variability across the various wavelengths follows the expectations of
models in which rather weak shock waves propagate down a jet containing
a turbulent magnetic field. After taking into account relativistic aberration,
we find that the shock fronts  generally lie at oblique angles to the jet axis. 

\begin{figure}
\epsscale{1.0}
\plotone{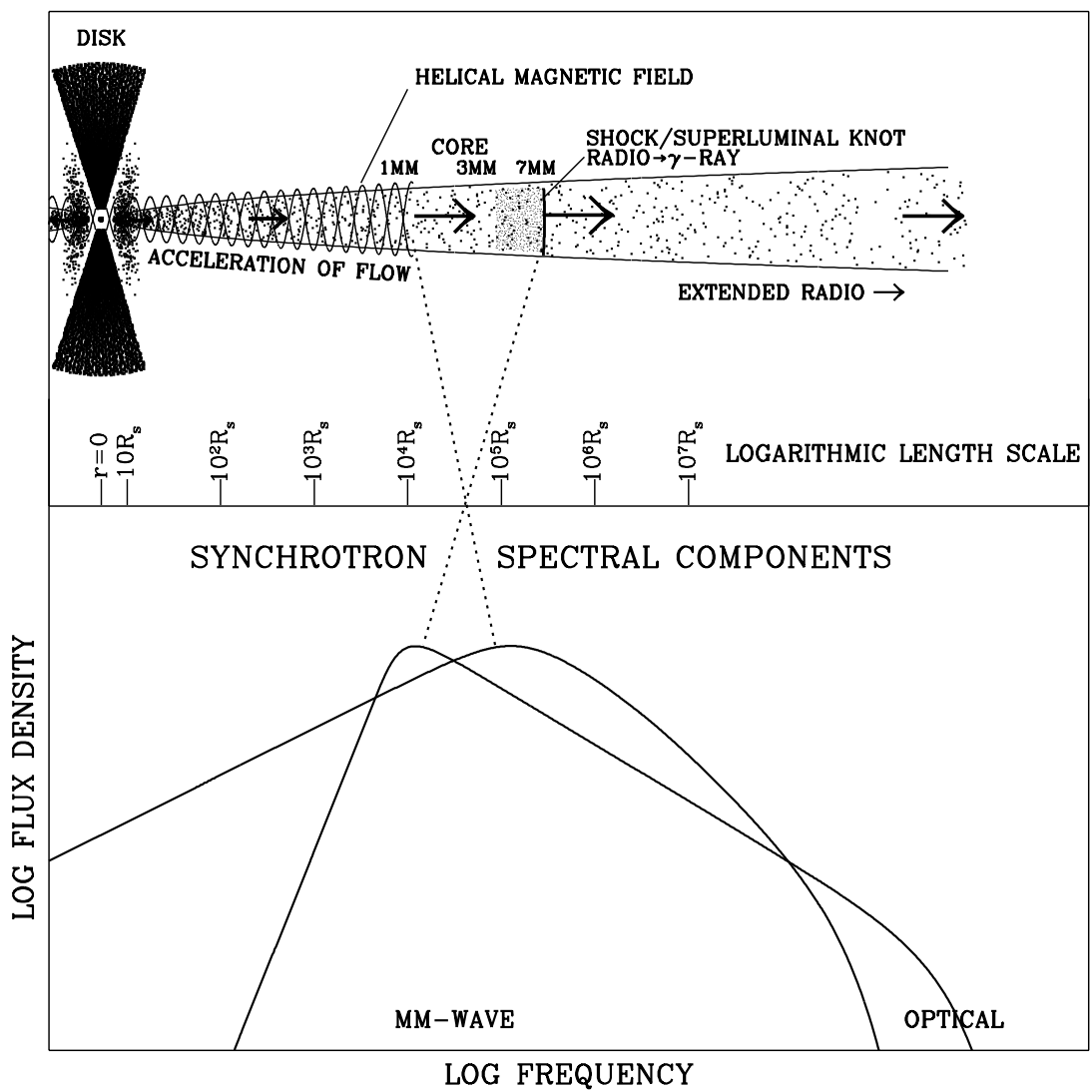}
\caption{Sketch of the main emission regions in the compact jet compatible with the results of our investigation. Note the logarithmic scale of distance from the black hole (denoted by the dot at the center of the flared accretion disk). The magnetic field is a tight helix up to the ``true '' core at 1~mm, beyond which it is chaotic. The system is symmetric with respect to the accretion disk 
(only one side is shown). The spectrum of the ambient jet rises gradually up to the peak frequency, at which the optical depth at the outer edge of the acceleration zone is of order unity, and cuts off at infrared frequencies. The spectrum of a shock moving through the core is more highly inverted below the peak and its high-frequency cutoff extends beyond the optical region. Although only one is shown, multiple shocks are often present at any given time.} \label{Sketch}
\end{figure}

The correlations that we have uncovered tightly link the emission at optical
and 3~mm wavelengths to the VLBI core at 7~mm. At 1~mm there is a 
possible anticorrelation between total flux density and degree of polarization, 
but not at 7~mm or 3~mm. This suggests the presence of a well-ordered
magnetic field in a region that is optically thick at $\lambda \gtrsim 1$~mm but
does not contain electrons with energies sufficiently high to radiate at
optical wavelengths. During outbursts, the 1~mm polarization is weakened by
the somewhat more weakly polarized emission from a shock moving through the downstream
portion of the jet where the ambient magnetic field is chaotic. We suggest that
the quasi-steady 1~mm emission arises near the end of the section where the
jet is accelerated by magnetic forces, a process that requires a tight helical field
geometry. The acceleration of electrons in this region is apparently inadequate
to produce optical synchrotron radiation - shocks might be required to do this.
Figure \ref{Sketch} presents a sketch of a relativistic jet to illustrate
our interpretation.
    
Our study demonstrates the utility of multiwaveband polarization monitoring of
blazars when combined with multi-epoch imaging of their jets with VLBI at
mm wavelengths. Further progress is possible through more intense time and
frequency coverage of a large number of objects.

\acknowledgments
This material is based on work supported by the National Science Foundation 
under grant no. AST-0406865. P.S. Smith acknowledges support from National
Aeronautics and Space Administration contract 1256424. T.V. Cawthorne and 
A.M. Stirling acknowledge support through research grants from the UK Particle Physics 
and Astronomy Research Council. J.L. G\'omez acknowledges
support from the Spanish Ministerio de Educaci\'on y Ciencia and the European Fund 
for Regional Development through grant AYA2004-08067-C03-03.
We thank Gary Schmidt for the maintenance and use of the Two-Holer Polarimeter.
The VLBA is a facility of the National Radio Astronomy
Observatory, operated by Associated Universities Inc. under cooperative 
agreement with the National Science Foundation. The James Clerk Maxwell Telescope 
is operated by The Joint Astronomy Centre on behalf of the Particle Physics 
and Astronomy Research Council of the United Kingdom, 
the Netherlands Organisation for Scientific Research, and the National 
Research Council of Canada.

\begin{deluxetable}{crrrrrrr}
\singlespace
\tablecolumns{8}
\tablecaption{\small\bf Optical Polarization Data \label{TAB_Opt}} 
\tabletypesize{\footnotesize}
\tablehead{
\colhead{Source}&\colhead{Epoch}&\colhead{$Filt_{\rm S}$}&\colhead{$Mag$}
&\colhead{$I_{\rm opt}$[mJy]}
&\colhead{$Filt_{\rm m}$}&\colhead{$m_{\rm opt}$[\%]}&\colhead{$\chi_{\rm opt}$[$^\circ$]} \\
\colhead{(1)}&\colhead{(2)}&\colhead{(3)}&\colhead{(4)}&\colhead{(5)}&\colhead{(6)}&
\colhead{(7)}&\colhead{(8)}}
\startdata
3C 66A&1999/02/12&&\nodata&\nodata&$W$&13.8$\pm$0.2&37.2$\pm$0.3 \\
&1999/02/13&$V$&14.70$\pm$0.02&4.81$\pm$0.08&$W$&13.3$\pm$0.2&45.2$\pm$0.4 \\
&1999/02/14&$V$&14.74$\pm$0.02&4.63$\pm$0.08&$W$&14.5$\pm$0.2&47.4$\pm$0.3 \\
&2000/09/27&$V$&14.78$\pm$0.02&4.46$\pm$0.08&$W$&29.7$\pm$0.2&17.2$\pm$0.2 \\
&2000/09/28&$V$&14.81$\pm$0.02&4.35$\pm$0.09&$W$&26.8$\pm$0.3&17.1$\pm$0.4 \\
&2000/09/29&&\nodata&\nodata&$W$&24.2$\pm$0.3&13.4$\pm$0.4 \\
&2000/11/30&$V$&14.40$\pm$0.02&6.31$\pm$0.11&$W$&23.7$\pm$0.1&25.4$\pm$0.1 \\
&2000/12/01&&\nodata&\nodata&$W$&22.8$\pm$0.2&25.1$\pm$0.2 \\
&2001/01/19&$V$&14.43$\pm$0.02&6.16$\pm$0.10&$W$&24.6$\pm$0.1&10.1$\pm$0.1 \\
3C 111&1999/02/13&V&18.95$\pm$0.12&0.10$\pm$0.02&$W$&3.1$\pm$0.8&126$\pm$4 \\
&2000/04/03&&\nodata&\nodata&$W$&0.8$\pm$0.8&106$\pm$5 \\
&2000/09/27&&\nodata&\nodata&$W$&1.5$\pm$0.6&104$\pm$3 \\
&2000/11/30&&\nodata&\nodata&$W$&1.9$\pm$0.5&131$\pm$3  \\
&2000/12/01&&\nodata&\nodata&$W$&1.6$\pm$0.8&110$\pm$5  \\
&2001/01/19&&\nodata&\nodata&$W$&1.3$\pm$0.4&89$\pm$3 \\
\enddata
\tablecomments{The table appears in full form electronically.}
\end{deluxetable}
\begin{deluxetable}{crrrrr}
\singlespace
\tablecolumns{6}
\tablecaption{\small\bf JCMT Polarization Data \label{TAB_JCMT}} 
\tabletypesize{\footnotesize}
\tablehead{
\colhead{Source}&\colhead{Epoch}&\colhead{$\lambda$[mm]}&\colhead{$I_{\rm 1mm}$[Jy]}
&\colhead{$m_{\rm 1mm}$[\%]}&\colhead{$\chi_{\rm 1mm}$[$^\circ$]}\\
\colhead{(1)}&\colhead{(2)}&\colhead{(3)}&\colhead{(4)}&\colhead{(5)}&\colhead{(6)}
}
\startdata
3C 66A&1998/05/15&1.35&0.38$\pm$0.04&36$\pm$4&24$\pm$2 \\
&1998/07/17&1.35&0.41$\pm$0.04&19$\pm$4&176$\pm$6 \\
&1998/09/30&1.35&0.45$\pm$0.04&14$\pm$2&56$\pm$4 \\
&1998/11/23&1.35&0.41$\pm$0.04&(11$\pm$5)&(172$\pm$10) \\
&1998/12/13&1.35&0.47$\pm$0.05&14$\pm$2&38$\pm$4 \\
&1999/02/18&1.35&0.43$\pm$0.04&11$\pm$2&157$\pm$4 \\
&1999/04/22&1.35&0.48$\pm$0.04&9.3$\pm$2.5&134$\pm$7 \\
&1999/09/23&1.35&0.59$\pm$0.06&9.7$\pm$2.5&158$\pm$10 \\
&2000/06/08&0.85&0.61$\pm$0.06&9.6$\pm$1.3&9$\pm$4 \\
&2001/01/17&0.85&0.61$\pm$0.06&11$\pm$2&23$\pm$4 \\
&2001/01/23&0.85&0.59$\pm$0.06&14$\pm$3&4$\pm$5 \\
3C 111&1998/07/17&1.35&1.03$\pm$0.10&12$\pm$2&39$\pm$4 \\
&1998/09/30&1.35&1.45$\pm$0.10& 3.9$\pm$0.7&117$\pm$5 \\
&1998/11/23&1.35&1.26$\pm$0.13&4.2$\pm$0.9&123$\pm$6 \\
&1998/12/13&1.35&1.37$\pm$0.14&5.8$\pm$1.6&39$\pm$7 \\
&1999/02/18&1.35&1.04$\pm$0.10&4.5$\pm$1.1&126$\pm$6 \\
&1999/09/23&1.35&0.89$\pm$0.09&6.4$\pm$2.0&143$\pm$11 \\
&1999/12/24&0.85&0.93$\pm$0.10&(4.9$\pm$1.9)&(17$\pm$10) \\
&2000/06/08&0.85&1.05$\pm$0.11&(1.0$\pm$1.5)&(144$\pm$39) \\
&2001/01/17&0.85&3.61$\pm$0.36&2.2$\pm$0.7&46$\pm$8 \\
&2001/01/23&0.85&5.99$\pm$0.60&1.0$\pm$0.3&17$\pm$7 \\
\enddata
\tablecomments{The table appears in full form electronically.}
\end{deluxetable}
\begin{deluxetable}{crrrr}
\singlespace
\tablecolumns{5}
\tablecaption{\small\bf BIMA Polarization Data \label{TAB_BIMA}}
\tabletypesize{\footnotesize}
\tablehead{
\colhead{Source}&\colhead{Epoch}&\colhead{$I_{\rm 3mm}$[Jy]}
&\colhead{$m_{\rm 3mm}$[\%]}&\colhead{$\chi_{\rm 3mm}$[$^\circ$]}\\
\colhead{(1)}&\colhead{(2)}&\colhead{(3)}&\colhead{(4)}&\colhead{(5)}}
\startdata
3C 66A  &2000/04/21& 0.82$\pm$0.01& 3.5$\pm$ 0.7&171$\pm$ 5 \\
&2000/04/25& 0.98$\pm$0.01& 5.0$\pm$ 0.5&  1$\pm$ 2 \\
&2001/02/06& 0.89$\pm$0.01& 9.3$\pm$ 0.3& 17$\pm$ 1 \\
&2001/03/20& 0.65$\pm$0.01& 4.2$\pm$ 0.2&  2$\pm$ 1 \\
&2001/03/29& 0.85$\pm$0.01& 5.9$\pm$ 0.6& 10$\pm$ 2 \\
&2001/04/03& 0.86$\pm$0.01& 5.0$\pm$ 0.5&  8$\pm$ 2 \\
&2001/04/07& 0.71$\pm$0.01& 4.8$\pm$ 0.4& 17$\pm$ 2 \\
&2001/04/12& 0.90$\pm$0.01& 3.3$\pm$ 0.8& 32$\pm$ 6 \\
&2001/04/16& 0.71$\pm$0.01& 3.1$\pm$ 0.6&  0$\pm$ 5 \\
&2001/04/20& 1.00$\pm$0.02& 5.6$\pm$ 0.8& 15$\pm$ 4 \\
3C 111  &2000/04/04& 2.63$\pm$0.01& 0.3$\pm$ 0.1&134$\pm$ 9 \\
&2000/04/14& 3.33$\pm$0.02& 2.0$\pm$ 1.2& 45$\pm$14 \\
&2000/12/06& 4.10$\pm$0.01& 0.5$\pm$ 0.2&138$\pm$11 \\
\enddata
\tablecomments{The table appears in full form electronically.}
\end{deluxetable}
\begin{deluxetable}{crrrrr}
\singlespace
\tablecolumns{6}
\tablecaption{\small\bf VLBA Polarization Data for the Core \label{TAB_VLBA}}
\tabletypesize{\footnotesize}
\tablehead{
\colhead{Source}&\colhead{Epoch}&\colhead{$I_{\rm 7mm}$[Jy]}&
\colhead{$\Theta_{\rm jet}$[$^\circ$]}
&\colhead{$m_{\rm 7mm}$[\%]}&\colhead{$\chi_{\rm 7mm}$[$^\circ$]}\\
\colhead{(1)}&\colhead{(2)}&\colhead{(3)}&\colhead{(4)}&\colhead{(5)}&\colhead{(6)}
}
\startdata
3C 66A&1998/03/25&0.52$\pm$0.05&$-$146$\pm$1&2.3$\pm$0.6&26$\pm$7 \\
&1998/05/30&0.45$\pm$0.07&$-$155$\pm$1&3.7$\pm$0.8&13$\pm$5 \\
&1998/07/31&0.30$\pm$0.03&$-$150.7$\pm$0.5&2.9$\pm$0.6&35$\pm$6 \\
&1998/10/05&0.31$\pm$0.04&$-$155$\pm$1&2.7$\pm$1.1&36$\pm$10\\
&1998/12/10&0.39$\pm$0.12&$-$159$\pm$2&3.1$\pm$1.3&13$\pm$11\\
&1999/02/11&0.48$\pm$0.05&$-$146$\pm$2&2.2$\pm$0.5&16$\pm$6 \\
&1999/04/29&0.50$\pm$0.03&$-$167$\pm$1&1.4$\pm$0.6&$-$6$\pm$11\\
&1999/07/18&0.47$\pm$0.02&$-$162$\pm$2&2.3$\pm$1.0&27$\pm$11\\
&1999/10/06&0.69$\pm$0.03&$-$154.4$\pm$0.5&3.2$\pm$0.4&18$\pm$4 \\
&1999/12/05&0.75$\pm$0.03&$-$159$\pm$1&7.7$\pm$0.4&5$\pm$3 \\
&2000/01/24&0.61$\pm$0.05&$-$155.4$\pm$0.5&3.9$\pm$0.4&0$\pm$3 \\
&2000/04/05&0.60$\pm$0.05&$-$158$\pm$1&3.8$\pm$0.5&14$\pm$4 \\
&2000/07/17&0.63$\pm$0.08&$-$163$\pm$1&4.3$\pm$1.7&28$\pm$11\\
&2000/10/01&0.73$\pm$0.04&$-$159$\pm$1&6.1$\pm$0.3&28$\pm$3 \\
&2000/12/11&0.60$\pm$0.06&$-$160$\pm$2&5.7$\pm$1.0&27$\pm$10\\
&2001/01/28&0.63$\pm$0.07&$-$162$\pm$2&6.1$\pm$1.5&35$\pm$8\\
&2001/04/14&0.60$\pm$0.05&$-$160$\pm$1&5.4$\pm$0.3&28$\pm$3\\
\enddata
\tablecomments{The table appears in full form electronically.}
\end{deluxetable}
\begin{deluxetable}{cccrrrrr}
\singlespace
\tablecolumns{8}
\tablecaption{\small\bf VLBA Polarization Data for the Jet Component \label{TAB_Pcomp}}
\tabletypesize{\footnotesize}
\tablehead{
\colhead{Source}&\colhead{Epoch}&\colhead{Comp.}&\colhead{$I^{\rm comp}_{\rm 7mm}$[Jy]}&
\colhead{$R$[mas]}&\colhead{$\Theta$[$^\circ$]}
&\colhead{$m^{\rm comp}_{\rm 7mm}$[\%]}&\colhead{$\chi^{\rm comp}_{\rm 7mm}$[$^\circ$]}
\\
\colhead{(1)}&\colhead{(2)}&\colhead{(3)}&\colhead{(4)}&\colhead{(5)}&\colhead{(6)}&
\colhead{(7)}&\colhead{(8)}
}
\startdata
3C 111&1998/03/25&$C1$&0.43$\pm$0.08&2.50$\pm$0.10&66.8$\pm$0.5&6.3$\pm$2.1&$-$44$\pm$10\\
&&$c1$&0.14$\pm$0.06&2.15$\pm$0.08&67.6$\pm$0.5&15.7$\pm$2.8&$-$43$\pm$8 \\
&1998/05/30&$C1$&0.54$\pm$0.08&2.71$\pm$0.10&67$\pm$1&6.8$\pm$1.8&$-$67$\pm$7 \\
&&$c1$&0.10$\pm$0.04&2.21$\pm$0.10&69$\pm$1&36$\pm$6&$-$65$\pm$6 \\
&1998/07/31&$C1$&0.32$\pm$0.08&2.93$\pm$0.10&67.7$\pm$0.5&9.4$\pm$2.1&$-$55$\pm$7\\
&&$c1$&0.16$\pm$0.05&2.34$\pm$0.15&68.5$\pm$0.5&13$\pm$3&$-$47$\pm$8 \\
&1998/10/05&$C1$&0.22$\pm$0.06&3.31$\pm$0.15&67.5$\pm$1.0&9.3$\pm$2.5&$-$45$\pm$8\\
&&$c1$&0.09$\pm$0.04&2.64$\pm$0.15&70$\pm$2&32$\pm$6&$-$45$\pm$6 \\
&1998/12/10&$C1$&0.17$\pm$0.05&3.73$\pm$0.15&67$\pm$1&15$\pm$4&$-$40$\pm$10 \\
&&$c1$&0.16$\pm$0.04&3.16$\pm$0.15&69$\pm$2&8.8$\pm$2.1&$-$54$\pm$10 \\
&1999/02/11&$C1$&0.17$\pm$0.05&4.01$\pm$0.15&67$\pm$1&18.8$\pm$4.5&$-$57$\pm$10 \\
&&$c1$&0.15$\pm$0.05&3.41$\pm$0.15&69$\pm$2&8.8$\pm$2.1&$-$60$\pm$10 \\
&1999/04/29&$C1$&0.12$\pm$0.05&4.33$\pm$0.20&66$\pm$2&21$\pm$5&$-$22$\pm$20 \\
&&$c1$&0.16$\pm$0.07&3.62$\pm$0.20&70$\pm$2&15.6$\pm$4.3&$-$41$\pm$20 \\
&1999/07/18&$C1$&0.20$\pm$0.05&4.69$\pm$0.20&66$\pm$2&6.0$\pm$1.5&$-$29$\pm$12 \\
&&$c2$&0.06$\pm$0.04&4.11$\pm$0.20&68$\pm$2&37$\pm$7&$-$31$\pm$20 \\
&&$c1$&0.04$\pm$0.03&3.69$\pm$0.20&70$\pm$2&38$\pm$8&$-$26$\pm$20 \\
&1999/10/06&$C1$&0.19$\pm$0.05&5.16$\pm$0.15&64$\pm$1&8.9$\pm$1.6&$-$27$\pm$8  \\
&&$c2$&0.07$\pm$0.03&4.63$\pm$0.15&67$\pm$1&20.0$\pm$4.5&$-$37$\pm$10 \\
&&$c1$&0.11$\pm$0.04&3.90$\pm$0.15&69$\pm$1&10.9$\pm$3.4&$-$31$\pm$10 \\
&1999/12/05&$C1$&0.12$\pm$0.05&5.46$\pm$0.15&64$\pm$2&14.5$\pm$3.6&$-$28$\pm$10 \\
&&$c2$&0.06$\pm$0.03&4.98$\pm$0.20&66$\pm$2&27$\pm$5&$-$43$\pm$15 \\
&&$c1$&0.10$\pm$0.04&4.11$\pm$0.20&68$\pm$2&18.0$\pm$4.0&$-$50$\pm$15 \\
&2000/04/05&$c2$&0.05$\pm$0.03&5.37$\pm$0.20&64$\pm$2&22$\pm$5&$-$23$\pm$20 \\
&&$c1$&0.09$\pm$0.04&4.52$\pm$0.20&64$\pm$2&15.6$\pm$4.5&$-$42$\pm$20 \\
&2000/07/17&$c2$&0.05$\pm$0.03&5.90$\pm$0.20&64$\pm$2&22.0$\pm$4.5&$-$23$\pm$20 \\
&&$c1$&0.09$\pm$0.04&4.69$\pm$0.20&64$\pm$2&13.3$\pm$4.0&$-$42$\pm$20 \\
\enddata
\tablecomments{The table appears in full form electronically.}
\end{deluxetable}

\begin{deluxetable}{lrrrrr}
\singlespace
\tablecolumns{6}
\tablecaption{\bf{Total and Polarized Flux Spectral Indices}  \label{TAB_Spec}}
\tabletypesize{\footnotesize}
\tablehead{
\colhead{Source}&\colhead{Group}&\colhead{$\alpha_{\rm opt/1mm}$}&\colhead{$\alpha^{\rm p}_{\rm opt/1mm}$}
&\colhead{$\alpha_{\rm mm}$}&\colhead{$\alpha^{\rm p}_{\rm mm}$}\\
\colhead{(1)}&\colhead{(2)}&\colhead{(3)}&\colhead{(4)}&\colhead{(5)}&\colhead{(6)}
}
\startdata
3C 111 &LVP&0.55$\pm$0.02&0.60$\pm$0.04&0.52$\pm$0.05&0.15$\pm$0.20 \\
3C 120 &LVP&0.56$\pm$0.02&1.03$\pm$0.05&0.47$\pm$0.02&$-$0.28$\pm$0.34 \\
3C 273&LVP&0.66$\pm$0.01&1.04$\pm$0.07&0.64$\pm$0.16&0.51$\pm$0.27  \\
3C 66A&IVP&0.60$\pm$0.01&0.50$\pm$0.02&0.16$\pm$0.09&$-$0.11$\pm$0.21 \\
3C 279&IVP&1.19$\pm$0.01&1.01$\pm$0.04&0.08$\pm$0.06&$-$0.13$\pm$0.03  \\
3C 345&IVP&1.06$\pm$0.01&0.95$\pm$0.01&0.53$\pm$0.06&0.06$\pm$0.36 \\
1803+784&IVP&\nodata&\nodata&0.28$\pm$0.17&0.19$\pm$0.15 \\         
1823+568&IVP&\nodata&\nodata&0.27$\pm$0.17&0.12$\pm$0.23   \\
BL Lac &IVP&0.50$\pm$0.01&0.54$\pm$0.02&0.15$\pm$0.04&0.08$\pm$0.09 \\
0420$-$014&HVP&1.00$\pm$0.01&0.86$\pm$0.01&0.25$\pm$0.03&0.11$\pm$0.05 \\
0528+134&HVP&0.96$\pm$0.02&0.98$\pm$0.05&0.48$\pm$0.03&$-$0.32$\pm$0.26 \\
OJ 287&HVP&0.90$\pm$0.01&0.80$\pm$0.01&0.24$\pm$0.06&$-$0.25$\pm$0.12 \\
1510$-$089&HVP&0.90$\pm$0.01&1.13$\pm$0.04&0.68$\pm$0.07&$-$0.44$\pm$0.17 \\
CTA 102&HVP&1.00$\pm$0.02&0.97$\pm$0.06&0.37$\pm$0.20&0.11$\pm$0.12  \\
3C 454.3&HVP&0.97$\pm$0.01&1.05$\pm$0.02&0.40$\pm$0.13&$-$0.11$\pm$0.32  \\ 
\enddata
\tablecomments{uncertainties for $\alpha_{\rm opt/1mm}$ are determined from uncertainties
of flux measurements; uncertainties for $\alpha_{\rm mm}$ are based on consistency between
different wavelengths}
\end{deluxetable}

\begin{deluxetable}{llrrrrrrl}
\singlespace
\tablecolumns{9}
\tablecaption{\bf Rotation Measures \label{TAB_RM}}
\tabletypesize{\footnotesize}
\tablehead{
\colhead{Source}&\colhead{z}&\colhead{Group}&\colhead{$\Delta\chi_{\rm 3mm}$}&\colhead{$\Delta\chi_{\rm 7mm}$}
&\colhead{$RM^\circ_{\rm 43GHz}$}&\colhead{$RM^\circ_{\rm 15GHz}$}&\colhead{$RM^\circ_{\rm 8GHz}$}&\colhead{$a$} \\
\colhead{}&\colhead{}&\colhead{}&\colhead{deg.}&\colhead{deg.}&\colhead{10$^3$rad~m${-2}$}&
\colhead{10$^3$rad~m${-2}$}&\colhead{10$^3$rad~m${-2}$}&\colhead{} \\
\colhead{(1)}&\colhead{(2)}&\colhead{(3)}&\colhead{(4)}&\colhead{(5)}&\colhead{(6)}&
\colhead{(7)}&\colhead{(8)}&\colhead{(9)}
}
\startdata
3C 66A &0.444&IVP&$-$3&$-$16&12$\pm$6&\nodata&\nodata&\nodata\\
3C 279&0.538&IVP&$-$3&$-$16&14$\pm$2&$-$5.7 (2)& 3.1 (1)&0.92$\pm$0.04     \\
3C 345&0.595&IVP&$-$2&$-$9&11$\pm$7&\nodata&$-$0.33$^*$ (3)&1.6 \\
1803+784&0.68&IVP&$-$3&$-$16&12$\pm$9&\nodata&0.56 (1)&1.8   \\           
1823+568&0.664&IVP&1&7&$-$10$\pm$1&1.1 (2)& 0.36 (1)&1.81$\pm$0.02 \\
BL Lac&0.069&IVP&$-$3&$-$16&6.7$\pm$5.5&7.0 (2)& 0.42 (1)&1.5$\pm$1.2 \\
0420$-$014&0.915&HVP&4&19&$-$24$\pm$3&\nodata&\nodata&\nodata\\
0528+134&2.06&HVP&$-$16&$-$87&280$\pm$30&\nodata&1.5 (1)&3.1 \\
OJ 287&0.306&HVP&6&30&$-$19$\pm$11&\nodata&\nodata&\nodata \\
1510$-$089&0.361&HVP&$-$7&$-$38&26$\pm$10&\nodata&\nodata&\nodata \\
CTA 102&1.037&HVP&18&97&$-$140$\pm$40&\nodata&$-$4.2$^*$ (4)&1.6  \\
3C 454.3&0.859&HVP&$-4$&$-19$&24$\pm$13&\nodata&9.0 (1)&1.8 \\
\enddata
\tablecomments{Columns are as follows: (1) - name of the source; (2) - redshift; 
(3) - polarization variability group; (4) - $RM$ correction
at 3~mm; (5)- $RM$ correction at 7~mm, (6) - intrinsic rotation
measure obtained in the 43~GHz core;
(7) - intrinsic rotation measure obtained in the 15~GHz core; 
(8) - intrinsic rotation measure obtained in the 8~GHz core; 
$^*$ - intrinsic rotation measure obtained in the core at 5~GHz instead of 8~GHz; 
numbers in parentheses denote references for $RM^{\circ}$ measured at low
frequencies: 
1 - \citet{ZT03}, 2 - \citet{GRSS06}, 3 - \citet{TAYLOR98}, 4 - \citet{TAYLOR00};
(9) - the exponent in the relation $RM\propto~d^{-a}$.}  
\end{deluxetable}

\begin{deluxetable}{lrrrrrr}
\singlespace
\tablecolumns{7}
\tablecaption{\bf Correlation Coefficients between Polarization in
the Core at 7~mm and Overall Polarization at Optical, 1~mm, and 3~mm 
Wavelengths \label{TAB_Pcor}}
\tabletypesize{\footnotesize}
\tablehead{
\colhead{Source}&\colhead{$f^p_{o7}$}&\colhead{$f^p_{17}$}&\colhead{$f^p_{37}$}
&\colhead{$r^m_{o7}$}&\colhead{$r^m_{17}$}&\colhead{$r^m_{37}$} \\
\colhead{(1)}&\colhead{(2)}&\colhead{(3)}&\colhead{(4)}&\colhead{(5)}&\colhead{(6)}&
\colhead{(7)}
}
\startdata
3C 66A &   {\bf 0.81 (4)}&0.22 ( 8)&0.81 (3)&{\bf 0.97 (4)} &  0.21 (8) & 0.81 (3) \\
3C 111 &  \nodata (1)&\nodata (0)&\nodata (2)&\nodata (0)&\nodata ( 0) &\nodata  (0) \\
0420$-$014 & 0.67 (4)&{\bf 0.94 ( 5)}&\nodata (1)& 0.72 (5)&   0.55 ( 4) & \nodata (1) \\    
3C 120     & \nodata (1)&\nodata ( 2) &\nodata (1)&\nodata (1)&\nodata ( 2) &\nodata(1) \\   
0528+134   &\nodata (1)&$-$0.17 ( 8)&$-$0.38 ( 3) &\nodata (1)&$-$0.49 ( 8)&  0.14 (3) \\ 
OJ 287     &0.33 (6)&{\bf 0.78 ( 7)}&$-$0.11 (5)& 0.30 (6)&{\bf 0.68 ( 7)}&  0.25 (5) \\  
3C 273     &\nodata (1)&\nodata ( 2)&\nodata (2)&\nodata (1)&\nodata ( 2)&\nodata (2) \\
3C 279     &0.31 (5)&0.34 ( 9)&{\bf 0.85 (6)}&0.21 (5)&   0.39 ( 9)&{\bf 0.90 (6)} \\   
1510$-$089 &\nodata (1)& 0.79 ( 5)&{\bf 0.97 (4)}& 0.08 (3)&   0.65 ( 5)&{\bf 0.92 (4)} \\ 
3C 345     &$-$0.19 (4)&   0.30 ( 6)& 0.59 (5)&$-$0.03 (5)&{\bf 0.79 ( 6)}&   0.36 (5) \\
1803+784   &\nodata (0)&$-$0.09 ( 5)&$-$0.09 (5) &\nodata (0)&$-$0.70 ( 5)& 0.61 (5) \\
1823+568   &\nodata (0)&0.29 ( 5)&0.76 (4) & $-$0.89 (3)&   0.30 ( 6)&{\bf 0.94 (4)}\\
BL Lac     &{\bf $-$0.97 (3)} &$-$0.16 (12)&{\bf 0.98 (3)} &0.18 (4) &0.23 (12)&{\bf 0.81 (4)}\\
CTA 102    &$-$0.68 (3) &  0.34 (11)&{\bf 0.90 (4)} & $-$0.78 (3)& 0.18 (12)&0.72 (4) \\
3C 454.3   &\nodata (2)&$-$0.17 (11)&$-$0.14 (3) &$-$0.41 (4)&$-$0.14 (11)&$-$0.59 (3) \\     
\enddata
\tablecomments{In this and subsequent tables, coefficients of correlation indicated by bold font are significant at a significance level $\epsilon$=0.1; integers in parentheses indicate number of observations.}  
\end{deluxetable}

\begin{deluxetable}{lrrrr}
\singlespace
\tablecolumns{5}
\tablecaption{\bf Correlation Coefficients between Polarization in
the Inner Jet at 7~mm and Overall Polarization at 1~mm and 3~mm \label{TAB_PCcor}}
\tabletypesize{\footnotesize}
\tablehead{
\colhead{Source}&\colhead{${f^p}'_{17}$}&\colhead{${f^p}'_{37}$}&\colhead{${r^m}'_{17}$}&\colhead{${r^m}'_{37}$}\\
\colhead{(1)}&\colhead{(2)}&\colhead{(3)}&\colhead{(4)}&\colhead{(5)}
}
\startdata
3C 111 &0.25 ( 7)& \nodata  (2)&{\bf  0.64 ( 7)}& \nodata  (2) \\
3C 120 &{\bf 0.87 ( 5)}& \nodata (1)&{\bf  0.78 ( 6)}& \nodata  (1) \\  
3C 273  &{\bf 0.79 ( 8)}&{\bf 0.98 (5)}& 0.41 ( 9)&{\bf 0.99 (5)} \\          
CTA 102 &{\bf 0.70 (12)}& 0.51 (4)&{\bf 0.53 (12)}&{\bf 0.80 (4)} \\
3C 454.3 &0.36 (13)&0.22 (3)&{\bf 0.61 (13)}& 0.12 (3)\\
\enddata
\end{deluxetable}

\begin{deluxetable}{lrrrr}
\singlespace
\tablecolumns{5}
\tablecaption{\bf Correlation Coefficients between Degree of Polarization
and Total Flux Density \label{TAB_PTcor}}
\tabletypesize{\footnotesize}
\tablehead{
\colhead{Source}&\colhead{$r_{\rm opt}$}&\colhead{$r_{\rm 1mm}$}&\colhead{$r_{\rm 3mm}$}&
\colhead{$r_{\rm 7mm}$}\\
\colhead{(1)}&\colhead{(2)}&\colhead{(3)}&\colhead{(4)}&\colhead{(5)}
}
\startdata
3C 66A &0.08 (6) &$-$0.44 (11) & 0.37 (10)&{\bf 0.67 (17)} \\
3C 111 &\nodata(1)&$-$0.51 (10) & 0.06 ( 3)&\nodata ( 0) \\
0420$-$014 &{\bf 0.89 (5)}&$-$0.06 (11) &\nodata ( 1)&{\bf 0.43 (17)} \\ 
3C 120     &\nodata (2)&$-$0.48 ( 8) &\nodata ( 2)&\nodata ( 2) \\   
0528+134   &\nodata  (1)& $-$0.33 (11)&$-$0.89 ( 3) & 0.02 (17) \\
OJ 287     & 0.10 (8) &0.02 (17)&$-$0.36 (13)&$-$0.28 (17) \\
3C 273     &$-$0.01 (6)&0.20 (20)& 0.32 (20)&$-$0.06 ( 5)  \\
3C 279     &$-$0.13 (7)& 0.23 (22)&$-$0.13 (18)& 0.13 (17) \\
1510$-$089 &\nodata (2)& $-$0.62 ( 5)&{\bf 0.74 ( 9)}&$-$0.18 (17) \\
3C 345     &{\bf 0.98 (5)}& $-$0.41 (16)&0.34 (22)&$-$0.23 (17) \\
1803+784   &\nodata  (0)& 0.11 ( 6)&$-$0.35 (20)&{\bf 0.48 (17)} \\           
1823+568   &\nodata  (0)& 0.00 ( 8)&{\bf 0.43 (20)}& 0.38 (17) \\
BL Lac    &$-$0.57 (5)&{\bf $-$0.51 (19)}&$-$0.07 ( 6)&$-$0.12 (17) \\
CTA 102    &$-$0.43 (3)&    0.25 (13)&   0.36 ( 5)&0.20 (17) \\
3C 454.3   &0.44 (3)&    0.14 (14)&$-$0.19 ( 5)&{\bf 0.67 (17)} \\    
\enddata
\end{deluxetable}

\begin{deluxetable}{lrrrr}
\singlespace
\tablecolumns{5}
\tablecaption{\bf Correlation Coefficients between Polarized
and Total Flux Density \label{TAB_PPTcor}}
\tabletypesize{\footnotesize}
\tablehead{
\colhead{Source}&\colhead{$f_{\rm opt}$}&\colhead{$f_{\rm 1mm}$}&\colhead{$f_{\rm 3mm}$}&
\colhead{$f_{\rm 7mm}$}\\
\colhead{(1)}&\colhead{(2)}&\colhead{(3)}&\colhead{(4)}&\colhead{(5)}
}
\startdata
3C 66A &    0.61 (6) &0.12 (11) &{\bf 0.60 (10)}&{\bf 0.82 (17)} \\
3C 111 &       \nodata (1)&0.24 (10) & 0.15 ( 3)&\nodata ( 0) \\
0420$-$014 &{\bf 0.92 (5)}&{\bf 0.88 (11)} &\nodata ( 1)&{\bf 0.84 (17)} \\ 
3C 120     & \nodata (2)&0.05 ( 8) &\nodata ( 2)&\nodata ( 2) \\   
0528+134   &\nodata  (1)&0.32 (11)&$-$0.81 ( 3) &{\bf 0.55 (17)} \\
OJ 287     &{\bf 0.72 (8)}&{\bf 0.44 (17)}& 0.16 (13)&{\bf 0.60 (17)} \\
3C 273     & 0.21 (6)&{\bf 0.76 (20)}&{\bf 0.57 (20)}& 0.49 ( 5) \\
3C 279     &{\bf 0.78 (7)}&{\bf 0.60 (22)}&{\bf 0.61 (18)}&{\bf 0.69 (17)} \\
1510$-$089 &\nodata (2)& 0.03 ( 5)&{\bf 0.81 ( 9)}& 0.06 (17) \\
3C 345     &{\bf 0.99 (5)}& 0.21 (16)&{\bf 0.74 (22)}& 0.39 (17) \\
1803+784   &\nodata  (0)& 0.26 ( 6)& $-$0.20 (20)&{\bf 0.66 (17)} \\           
1823+568   &\nodata  (0)&{\bf 0.77 ( 8)}&{\bf 0.72 (20)} &{\bf 0.79 (17)} \\
BL Lac    & $-$0.14 (5)& 0.30 (19)&0.67 ( 6) &{\bf 0.66 (17)} \\
CTA 102    &$-$0.30 (3)&{\bf 0.60 (13)}& 0.52 ( 5)&{\bf 0.76 (17)} \\
3C 454.3   &0.63 (3)&{\bf 0.80 (20)}& 0.08 ( 5)&{\bf 0.91 (17)} \\    
\enddata
\end{deluxetable}

\begin{deluxetable}{llrrrr}
\singlespace
\tablecolumns{6}
\tablecaption{\bf Average Values of Fractional Polarization
\label{TAB_POL}}
\tabletypesize{\footnotesize}
\tablehead{
\colhead{Group}&\colhead{}&\colhead{Opt}&\colhead{1 mm}&\colhead{3 mm}&\colhead{7 mm}
}
\startdata
LVP &$\langle{m}_{\rm max}\rangle$[\%]&1.4$\pm$1.2 (1.0)&10.2$\pm$2.1 (7.2)&2.8$\pm$0.9 (2.0)&1.4$\pm$0.3 (1.0) \\ 
&$N^{\rm max}_{\rm cell}$&3104&47&625&2500 \\
HVP &$\langle{m}_{\rm max}\rangle$[\%]&11.4$\pm$7.5 (8.1)&8.6$\pm$1.8 (6.1)&3.1$\pm$1.3 (2.2)&6.0$\pm$1.9 (4.2) \\ 
&$N^{\rm max}_{\rm cell}$&47&66&510&136 \\
IVP &$\langle{m}_{\rm max}\rangle$[\%]&29.0$\pm$11.3 (20)&16.8$\pm$9.4 (11)&9.1$\pm$2.4 (6.4)&7.5$\pm$2.1 (5.3) \\ 
&$N^{\rm max}_{\rm cell}$&8&18&59&87 \\
LVP&$\langle{m}_{\rm min}\rangle$[\%]&0.4$\pm$0.3 (0.3)&1.3$\pm$0.3 (0.9)&0.9$\pm$0.6 (0.6)&0.1$\pm$0.1 (0.1) \\ 
&$N^{\rm min}_{\rm cell}$&38025&2900&6050&490000 \\
HVP&$\langle{m}_{\rm min}\rangle$[\%]&1.7$\pm$1.2 (1.2)&1.6$\pm$0.8 (1.1)&0.6$\pm$0.3 (0.4)&0.5$\pm$0.2 (0.3) \\ 
&$N^{\rm min}_{\rm cell}$&2105&1914&13611&19600 \\
IVP&$\langle{m}_{\rm min}\rangle$[\%]&10.6$\pm$8.2 (7.5)&4.0$\pm$3.1 (2.8)&3.2$\pm$2.1 (2.3)&2.0$\pm$1.0 (1.4) \\ 
&$N^{\rm min}_{\rm cell}$&54&306&480&1225 \\
\enddata
\tablecomments{Values in parentheses indicate standard deviations expected from the cellular model with number of
cells derived from the fractional polarization; see text} 
\end{deluxetable}

\begin{deluxetable}{lrrrrrrrrr}
\singlespace
\tablecolumns{10}
\tablecaption{\bf Derived Shock Strength \label{TAB_Shock}}
\tabletypesize{\footnotesize}
\tablehead{
\colhead{Source}&\colhead{$m^{\rm max}_{\rm opt}$[\%]}&\colhead{$\eta_{\rm opt}$}&
\colhead{$m^{\rm max}_{\rm 1mm}$[\%]}&\colhead{$\eta_{\rm 1mm}$}&
\colhead{$m^{\rm max}_{\rm 3mm}$[\%]}&\colhead{$\eta_{\rm 3mm}$}&
\colhead{$m^{\rm max}_{\rm 7mm}$[\%]}&\colhead{$\eta_{\rm 7mm}$}&
\colhead{$\Theta'$[$^\circ$]}\\
\colhead{(1)}&\colhead{(2)}&\colhead{(3)}&\colhead{(4)}&\colhead{(5)}&\colhead{(6)}&
\colhead{(7)}&\colhead{(8)}&\colhead{(9)}&\colhead{(10)}
}
\startdata
3C 66A &  29.7 & 1.94 &  36.4 & 3.20 &   9.3 &   1.21 &   7.7 &   1.17 & 116.0 \\
3C 111 &   3.1 &   1.05 &  12.2 &   1.22 &   2.3 &   1.04 &   0.5 &   1.01 & 108.3 \\
0420$-$014  &  26.2 &   1.81 &   7.5 &   1.18 &   0.5 &   1.01 &   3.3 &   1.07 &  59.8 \\ 
3C 120 &   0.5 &   1.01 &  11.1 &   1.30 &   2.1 &   1.05 &   1.1 &   1.02 & 124.5 \\
0528+134  &   7.0 &   1.15 &  12.1 &   1.32 &   1.4 &   1.03 &   5.4 &   1.12 &  57.0 \\
OJ 287 &  15.6 &   1.24 &   6.9 &   1.11 &   4.0 &   1.06 &   5.0 &   1.08 &  85.3 \\
3C 273 &   0.5 &   1.01 &   7.2 &   1.11 &   4.0 &   1.06 &   1.6 &   1.02 &  96.8 \\
3C 279 &  39.2 &   3.49 &  10.9 &   1.30 &  13.7 &   1.40 &   6.9 &   1.17 &  59.2 \\
1510$-$089  &   4.1 &   1.06 &   9.7 &   1.15 &   4.8 &   1.07 &   9.7 &   1.15 &  83.6 \\
3C 345 &  38.3 &   1.78 &  11.4 &   1.18 &   8.1 &   1.13 &   9.6 &   1.15 &  82.7 \\
1803+784  & \nodata & \nodata &  11.3 &   1.24 &   5.9 &   1.11 &   5.7 &   1.11 &  67.1\\
1823+568  &  29.8 & $--$ &  20.3 &$--$&   9.9 &   1.91 &  10.6 &   2.08 &  37.0 \\
BL Lac &   7.8 &   1.12 &  10.4 &   1.18 &   7.8 &   1.13 &   4.9 &   1.08 &  86.3\\
CTA 102&   9.4 &   1.14 &   7.1 &   1.12 &   3.4 &   1.06 &   6.1 &   1.10 &  75.9 \\
3C 454.3 &   6.2 &   1.28 &   8.5 &   1.51 &   1.8 &   1.07 &   6.6 &   1.35 &  39.0 \\
\enddata
\end{deluxetable}

\clearpage

\clearpage
\clearpage

\clearpage
\end{document}